\documentclass[journal,twocolumn,10pt]{IEEEtran}
\usepackage{multirow}
\usepackage{textcomp}
\usepackage{stfloats}
\usepackage{url}
\usepackage{graphicx}
\usepackage{verbatim}
\usepackage{upgreek}
\usepackage{amssymb,amsmath}
\usepackage{color}
\usepackage{epstopdf}
\usepackage{bm}
\usepackage{cite}
\usepackage{array,color}
\usepackage{algorithm}
\usepackage{algpseudocode}
\usepackage{amsmath}
\usepackage{amsfonts}
\usepackage{amsthm}
\usepackage{graphics}
\usepackage{epsfig}
\usepackage{soul}
\usepackage{gensymb}
\soulregister\cite7
\soulregister\ref7
\usepackage{subfigure}
\usepackage{setspace}
\usepackage{makecell}

\usepackage{amsmath}

\begin{document}

\title{Annealed Interference Cancellation for AFDM Symbol Detection: Temperature-Dependent Convergence Analysis and Parameter Design}

\author{Shuntian~Tang, Zesong~Fei,~\IEEEmembership{Senior~Member,~IEEE}, Xinyi~Wang,~\IEEEmembership{Member,~IEEE}, Zhiqiang~Wei,~\IEEEmembership{Member,~IEEE}, Shuangyang~Li,~\IEEEmembership{Member,~IEEE}, Tongyang~Xu,~\IEEEmembership{Senior~Member,~IEEE}

\thanks{A part of this paper has been submitted to the 2026 IEEE Global Communications Conference (GLOBECOM).}

\thanks{Shuntian Tang, Zesong Fei, and Xinyi Wang are with the School of Information and Electronics, Beijing Institute of Technology, Beijing 100081, China (E-mail:	tst491961069@163.com, feizesong@bit.edu.cn, wangxinyi@bit.edu.cn).}

\thanks{Zhiqiang Wei is with the School of Mathematics and Statistics, Xi’an Jiaotong University, Xi’an 710049, China, and also with the Peng Cheng Laboratory, Shenzhen, Guangdong 518055, China, and also with the Pazhou Laboratory (Huangpu), Guangzhou, Guangdong 510555, China (email: zhiqiang.wei@xjtu.edu.cn).}

\thanks{Shuangyang Li is with the Department of Electrical Engineering and Computer Science, Technical University of Berlin, 10587 Berlin, Germany (e-mail: shuangyang.li@tu-berlin.de).}

\thanks{Tongyang Xu is with the Department of Electronic and Electrical Engineering, University College London, WC1E 7JE London, U.K.
(e-mail: tongyang.xu.11@ucl.ac.uk).}


}



\maketitle

\begin{abstract}
	Low-complexity and reliable symbol detection remains a key challenge for
	affine frequency division multiplexing (AFDM) over doubly dispersive
	channels. In this paper, we develop a low-complexity detection framework
	based on the Ungerboeck observation model. By exploiting the energy
	concentration of the matched-filtered channel matrix, an
	interference-cancellation (IC) detector is employed for efficient symbol
	detection. To characterize its performance degradation in high signal-to-noise-ritio (SNR), a
	nonlinear error-evolution model is established, showing that the detector
	eventually enters either a fixed point or a finite-period limit cycle under
	the high-SNR approximation. Motivated by this phenomenon, we propose an
	annealed IC detector that incorporates temperature-controlled posterior
	updates and damping to avoid undesirable attractors. A
	temperature-dependent potential function and time-varying Jacobian
	analysis are further developed to guide the selection of the initial
	temperature and cooling factor. Simulation results demonstrate that the
	proposed detector effectively eliminates the high-SNR error floor,
	outperforms other symbol detectors, and
	maintains substantially lower computational complexity.
\end{abstract}


\section{Introduction}

\IEEEPARstart{T}{he} next generation of wireless communication systems is
expected to support highly reliable transmission in high-mobility
scenarios, including unmanned aerial
vehicles (UAVs), high-speed trains, and low-Earth-orbit (LEO) satellite systems
\cite{Saad2020NET,Liu2025TWC,Chai2022TSP, Seong2025TSP}. The combination of high mobility and high
carrier frequencies can induce substantial Doppler shifts, posing
significant challenges to reliable signal transmission. Although
orthogonal frequency division multiplexing (OFDM) has been widely adopted
as a conventional multicarrier waveform, its performance degrades severely
under large Doppler shifts because the resulting inter-carrier interference (ICI)
destroys subcarrier orthogonality~\cite{Qiu2019WC}.

To overcome this limitation, new modulation schemes, including orthogonal time frequency space (OTFS)~\cite{Deng2025CST, Shen2019TSP}, orthogonal chirp division multiplexing (OCDM)~\cite{Omar2021TWC}, and affine frequency division multiplexing (AFDM)~\cite{Bemani2023TWC}, have emerged as promising solutions owing to their inherent robustness against Doppler effects.
OTFS represents a doubly selective channel in the delay-Doppler (DD)
domain, where the rapidly time-varying channel exhibits an approximately
two-dimensional (2D) quasi-static structure~\cite{Zhang2026TCom}. OCDM, based
on the discrete Fresnel transform (DFnT), modulates information symbols
onto a set of orthogonal chirp subcarriers~\cite{Wang2021WCL}. By contrast,
AFDM employs the inverse discrete affine Fourier transform (IDAFT) with
adjustable chirp parameters, providing additional degrees of freedom for
signal representation in the affine-frequency (AF) domain and enabling
effective separation of different delay--Doppler channel components.
Although these waveforms generally achieve superior error-rate performance
over OFDM in doubly selective channels
\cite{Aslandogan2026TVT,Omar2022CL,Sui2025arXiv}, they exhibit different
implementation characteristics. The 2D signal structure of
OTFS may require substantial modifications to conventional OFDM
transceiver processing. Meanwhile, the performance of OCDM can be sensitive
to the mismatch between its chirp rate and the specific delay-Doppler
profile of the channel~\cite{Bemani2023TWC}. In comparison, AFDM retains a
one-dimensional (1D) implementation structure similar to OFDM, while allowing its chirp parameters to be adapted to diverse channel conditions. 

Owing to its implementation compatibility and waveform flexibility, AFDM
has attracted growing research interest as a promising solution for
high-mobility communications. In~\cite{Li2025TCom}, the effects of AFDM
parameter configurations on communication performance under minimum mean
square error (MMSE) equalization were investigated, providing useful
guidelines for parameter selection and diversity enhancement. The effects
of hardware impairments on AFDM were studied
in~\cite{Sui2026TWC}, demonstrating its great robustness against carrier frequency offset (CFO) and phase noise (PN). To improve spectral efficiency, a non-orthogonal AFDM waveform was
proposed in~\cite{Yi2026TWC}, where a bandwidth-compression factor was
incorporated into the AFDM modulator to enable controllable subcarrier
overlap. Besides, the time and frequency offset estimations for AFDM synchronization have
been studied in~\cite{Wang2025TWC}.
Beyond communication-oriented designs, AFDM has also shown considerable
potential for integrated sensing and communication (ISAC)~\cite{Zhu2024WCL, Xie2026TSP, Luo2025IOT}. In
~\cite{Zhu2024WCL}, a bistatic sensing-assisted channel estimation scheme
was proposed by exploiting the inherent sensing capabilities of AFDM. In~\cite{Xie2026TSP}, the impact of clutter on sensing performance was investigated, providing valuable insights into the practical deployment of AFDM-based sensing systems.
Moreover, AFDM-based ISAC was extended to near-field (NF) scenarios
in~\cite{Luo2025IOT}, further broadening its range of potential
applications.
	
Although AFDM has demonstrated considerable potential in various
application scenarios, its relatively high data-detection complexity
remains a major challenge for terminals with limited computational
capabilities~\cite{Li2026IOT}. In~\cite{Bemani2022ICASSP}, a
low-complexity MMSE detector and a weighted maximum-ratio combining
(MRC)-based decision-feedback equalizer were developed by exploiting the
sparsity of the AF-domain channel matrix. Building on this sparse channel
structure, several message-passing (MP)-based detectors have subsequently
been proposed
\cite{Anoop2026WCNCW1,Anoop2026WCNCW2,Guo2025ICCCS}. Among these detectors,
factor nodes (FNs) combine the messages received from variable nodes (VNs)
using the sum-product algorithm (SPA), thereby obtaining the symbol
posterior distributions required for maximum a posteriori (MAP) detection.
Since exact SPA processing entails considerable computational overhead,
various approximation strategies have been developed to simplify the
message updates. For example, the aggregate interference terms were
approximated as Gaussian random variables in~\cite{Wu2025ICC}, allowing
their messages to be characterized by only the corresponding means and
variances. In~\cite{Yi2026WCNCW}, an enhanced expectation propagation (EP)
detector was proposed, which exploits time-domain channel sparsity by
performing iterative equalization in the time domain rather than directly
in the AF domain.
Beyond approximating the messages in factor-graph-based detectors,
interference cancellation (IC) provides an alternative approach to
low-complexity symbol detection. In~\cite{Conti2026TWC}, an IC-based
detector was developed for OTFS under the Ungerboeck observation model,
which avoids edge-wise message passing and thereby offers a promising
design perspective for low-complexity AFDM detection. However, its
algorithmic parameters are mainly selected empirically, while the
underlying convergence behavior, particularly the performance degradation
in the high-SNR regime, remains insufficiently understood. Motivated by
this low-complexity architecture and the above unresolved issues, we extend
Ungerboeck-model-based IC detection to AFDM and develop a dynamical-system
framework for convergence characterization and parameter design.

In this paper, we propose a low-complexity annealed
interference-cancellation symbol detection framework for AFDM systems based
on the Ungerboeck observation model. We first formulate AFDM symbol
detection under the matched-filtered observation model, whose energy
concentration characteristics enables efficient interference-cancellation processing.
We then investigate the high-signal-to-noise-ratio (SNR) performance degradation of the
conventional IC detector and establish a nonlinear error-evolution model
to reveal its fixed-point and limit-cycle behaviors.
To overcome this limitation, we introduce a temperature-controlled
annealing mechanism into the posterior update and incorporate damping to
improve convergence stability. Furthermore, we develop a
temperature-dependent potential function and a time-varying Jacobian
analysis to characterize fixed-point motion, and tracking
behavior. Based on this analysis, practical strategies are derived for
selecting the initial temperature and cooling factor.

The main contributions are summarized as follows:
\begin{itemize}
	\item[$\bullet$] We introduce the Ungerboeck observation model for AFDM
	symbol detection, which concentrates the desired signal energy into the
	diagonal entries of the matched-filtered channel matrix and enables
	low-complexity interference-cancellation processing. We further investigate
	the high-SNR performance degradation of the conventional IC detector. Based on a finite-state nonlinear error-evolution
	model, we prove that the detector eventually enters either a fixed point
	or a finite-period limit cycle, thereby explaining its high-SNR error
	floor.
	
	\item[$\bullet$] We propose an annealed IC detector based on the
	Ungerboeck observation model. A temperature-controlled posterior update
	is introduced to preserve soft
	exploration during the early iterations and prevent premature
	convergence to undesirable attractors. Combined with a damping
	mechanism, the proposed detector effectively suppresses oscillatory
	behavior and eliminates the high-SNR error floor.
	
	\item[$\bullet$] We develop a temperature-dependent fixed-point and
	dynamical-system analysis to guide the annealing-parameter design. By
	constructing a potential function, analyzing its Hessian, and deriving
	the time-varying Jacobian, we characterize the fixed-point motion and tracking stability during annealing. Based on these results,
	an eigenvalue-based critical initial temperature
	$T_{\rm c}=-\lambda_{\min}(\mathbf{S})/(2\sigma_n^2)$ and a practical
	cooling-factor design principle are obtained. Simulation results verify
	the theoretical analysis and demonstrate superior BER performance with
	substantially lower computational complexity than representative
	benchmark detectors.
\end{itemize}

The remainder of this paper is organized as follows. 
Section~\ref{sec:system_model} introduces the AFDM system model and
the Ungerboeck observation model.
Section~\ref{Detection Algorithm} presents the conventional IC detector,
analyzes its high-SNR limitation, and develops the proposed annealed IC
detection algorithm.
Section~\ref{sec:dynamics} investigates the nonlinear error dynamics,
temperature-dependent fixed points, and tracking behavior, based on which
the initial temperature and cooling factor are designed.
Section~\ref{Simulation Results} presents the simulation results and
performance comparisons, and Section~\ref{Conclusion} concludes the paper.

\emph{Notations}: Unless otherwise specified, we use a lowercase letter $a$, a boldface lowercase letter $\mathbf{a}$, a boldface capital letter $\mathbf{A}$, and a calligraphy letter $\mathcal{A}$ to denote a scalar, a vector, a matrix, and a set, respectively. $\operatorname{diag}\left(\cdot\right)$ denotes the diagonal operation; $\left(\cdot\right)^*$, $\left(\cdot\right)^T$, and $\left(\cdot\right)^H$ denote the conjugate,  transpose, and Hermitian transpose operations, respectively; $\mathbb{R}$ and $\mathbb{C}$ denote the real and complex number fields, respectively; $\operatorname{Re}\{\cdot\}$ and $\operatorname{Im}\{\cdot\}$ denote real and imaginary parts of its argument, respectively; $\mathbf{I}$ denotes the identity matrix.

\section{PRELIMINARIES}
\label{sec:system_model}

\subsection{AFDM System}

At the AFDM transmitter, a complex symbol vector $\mathbf{x}=[x_0,x_1,\ldots,x_{N-1}]^T\in\mathbb{C}^{N\times1}$, whose entries are drawn from an $M$-ary constellation alphabet $\mathcal{A}$, is converted into a time-domain waveform through an $N$-point IDAFT. Accordingly, its $n$-th sample can be written as
\begin{equation}
	s_n=\frac{1}{\sqrt{N}}\sum_{m=0}^{N-1}x_m\psi(m,n),
	\quad n=0,1,\ldots,N-1,
\end{equation}
where
\begin{equation}
	\psi(m,n)=e^{j2\pi\left(c_1n^2+\frac{mn}{N}+c_2m^2\right)}
\end{equation}
is the orthogonal chirp basis function, and $c_1$ and $c_2$ are two adjustable chirp parameters. Before transmission, a chirp-periodic prefix (CPP) containing $L$ samples is inserted to prevent inter-symbol interference~(ISI) induced by the channel delay spread. In particular, the prefix is constructed as
\begin{equation}
	s_n=s_{N+n}
	e^{
		j2\pi c_1\left(N^2+2Nn\right)
		},
	\quad n=-L,\ldots,-1.
\end{equation}
For an even-valued $N$, if $2Nc_1$ is an integer, the CPP coincides with a conventional cyclic prefix (CP).

We consider a doubly dispersive channel composed of $P$ propagation paths. The time-varying discrete impulse response at time index $n$ is modeled as
\begin{equation}
	g_n(l)
	=
	\sum_{i=0}^{P-1}
	h_i
	e^{-j\frac{2\pi}{N}\nu_i n}
	\delta(l-l_i),
\end{equation}
where $h_i$, $l_i$, and $\nu_i$ represent the complex channel coefficient, normalized delay, and normalized Doppler shift of the $i$-th path, respectively. After propagation through this channel, the received time-domain signal is expressed as
\begin{equation}
	r_n
	=
	\sum_{i=0}^{P-1}
	h_i s_{n-l_i}
	e^{j\frac{2\pi}{N}\nu_i n}
	+w_n,
\end{equation}
where $w_n\sim\mathcal{CN}(0,\sigma_n^2)$ denotes additive white Gaussian noise	(AWGN).

At the receiver, the CPP is first discarded, after which an $N$-point discrete affine Fourier transform (DAFT) is performed. Denoting the resulting AF-domain observation by $\mathbf{y}=[y_0,y_1,\ldots,y_{N-1}]^T$, its $m$-th entry is given by
\begin{equation}
	y_m
	=
	\frac{1}{\sqrt{N}}
	\sum_{n=0}^{N-1}
	r_n\psi^*(m,n).
\end{equation}
The resulting AF-domain input--output model can then be represented in matrix form as
\begin{align}\label{IO-relation}
	\mathbf{y}
	&=
	\sum_{i=0}^{P-1}
	h_i\mathbf{H}_i\mathbf{x}
	+\overline{\mathbf{w}}=
	\mathbf{H}_{\mathrm{eff}}\mathbf{x}
	+\overline{\mathbf{w}},
\end{align}
where
\begin{equation}
	\mathbf{H}_{\mathrm{eff}}
	=
	\sum_{i=0}^{P-1}h_i\mathbf{H}_i
\end{equation}
denotes the effective AF-domain channel matrix. Since both the DAFT and IDAFT are unitary transformations, the transformed noise satisfies $\overline{\mathbf{w}}\sim\mathcal{CN}(\mathbf{0},\sigma^2_n\mathbf{I})$.

The AF-domain contribution associated with the $i$-th propagation path is described by
\begin{align}
	\mathbf{H}_i
	={}&
	\mathbf{\Lambda}_{c_2}
	\mathbf{F}
	\mathbf{\Lambda}_{c_1}
	\mathbf{\Gamma}_i
	\mathbf{\Delta}_{f_i}
	\mathbf{\Pi}^{l_i}
	\mathbf{\Lambda}_{c_1}^{H}
	\mathbf{F}^{H}
	\mathbf{\Lambda}_{c_2}^{H},
\end{align}
where $\mathbf{F}$ is the normalized $N$-point DFT matrix,
\begin{equation}
	\mathbf{\Lambda}_{c}
	=
	\operatorname{diag}
	\left(
	1,e^{-j2\pi c},\ldots,
	e^{-j2\pi c(N-1)^2}
	\right)
\end{equation}
is the chirp-phase matrix, $\mathbf{\Delta}_{f_i}$ accounts for the Doppler shift, and $\mathbf{\Pi}$ denotes the cyclic delay operator. The CPP-induced phase matrix $\mathbf{\Gamma}_i$ is diagonal, with its elements given by
\begin{equation}
	\mathbf{\Gamma}_i(n,n)
	=
	\begin{cases}
		e^{-j2\pi c_1\left[N^2-2N(l_i-n)\right]},
		& n<l_i,\\
		1,
		& n\geq l_i.
	\end{cases}
\end{equation}
When $N$ is even and $2Nc_1$ takes an integer value, $\mathbf{\Gamma}_i=\mathbf{I}$, which is consistent with the CP-equivalent case discussed above.

The structure of $\mathbf{H}_i$ depends strongly on whether the normalized Doppler shift is integer or fractional. For an integer Doppler shift, the nonzero entry in the $p$-th row is located at a single shifted column, and hence
\begin{equation} \label{channel_struc}
	\mathbf{H}_i(p, q)= \begin{cases}e^{j \frac{2 \pi}{N}\left(N c_1 l_i^2-q l_i+N c_2\left(q^2-p^2\right)\right)} & q=\left(p+\operatorname{loc}_i\right)_N, \\ 0 & \text { otherwise },\end{cases}
\end{equation}
where $\operatorname{loc}_i
=
\alpha_i+2Nc_1l_i$
specifies the AF-domain location associated with the $i$-th path, and $\alpha_i$ is the integer part of $\nu_i$. By selecting $c_1\geq\frac{2\alpha_{\max}+1}{2N}$,
each row and column of $\mathbf{H}_{\mathrm{eff}}$ consequently contains $P$ nonzero entries~\cite{Bemani2023TWC}.

When $\nu_i$ contains a fractional component, the energy associated with one path is no longer restricted to a single AF-domain position. In this case, the $(p,q)$-th entry of $\mathbf{H}_i$ becomes
\begin{align}\label{gen_channel_struc}
	\mathbf{H}_i(p, q)&= \frac{1}{N}e^{j\frac{2\pi}{N}\left(Nc_1l_i^2-ql_i+Nc_2\left(q^2-p^2\right)\right)} \notag \\
	&\times \sum_{n=0}^{N-1}e^{-j\frac{2\pi}{N}\left(\left(p-q+\nu_i+2Nc_1l_i\right)n\right)}.
\end{align}

The finite geometric sum in (\ref{gen_channel_struc}) produces a Dirichlet-kernel-type energy profile centered around $q=(p+\operatorname{loc}_i)_N$. Therefore, the magnitude of $\mathbf{H}_i(p,q)$ decreases as $q$ moves away from this dominant location. Although fractional Doppler shifts destroy the exact sparsity obtained in (\ref{channel_struc}), most of the channel energy remains concentrated within a limited neighborhood of the corresponding path location. As a result, $\mathbf{H}_{\mathrm{eff}}$ generally retains an approximately sparse structure.

\subsection{Ungerboeck Observation Model}

The AF-domain input-output relationship in~(\ref{IO-relation}) follows the Forney observation model. Alternatively, applying a matched filter to the received vector yields a sufficient statistic on which the symbol detectors developed in the following section are based
\begin{align}\label{Ungerboeck model}
	\mathbf{z} =\mathbf{H}_{\text{eff}}^H\mathbf{y}= \mathbf{G}\mathbf{x}+\widetilde{\mathbf{w}},
\end{align}
where $\widetilde{\mathbf{w}} = \mathbf{H}_{\text{eff}}^H\overline{\mathbf{w}}\sim\mathcal{CN}(0,\sigma^2_n\mathbf{G})$ is the matched-filtered colored Gaussian noise. Since (\ref{Ungerboeck model}) is obtained through matched filtering, its resulting observation model is the Ungerboeck one~\cite{Ungerboeck1974TCom,Conti2026TWC}.

Substituting (\ref{IO-relation}) into (\ref{Ungerboeck model}) gives
\begin{align}\label{structure G}
	\mathbf{G} =\underbrace{\sum_{i=0}^{P-1}\vert h_i \vert^2\mathbf{I}}_{\text{Dominant-diagonal}} + \underbrace{\sum_{i=0}^{P-1}\sum_{j=0,j \ne i}^{P-1}h^*_ih_j \mathbf{H}^H_i\mathbf{H}_j}_{\text{Off-diagonal}}.
\end{align}
The first term in (\ref{structure G}) results from the self-correlation of each propagation path and forms the diagonal component of $\mathbf{G}$. The second term captures the pairwise correlations between distinct paths and is primarily responsible for the off-diagonal coupling among the transmitted symbols.

For integer normalized Doppler shifts, i.e., $\nu_i\in\mathbb{Z}$, (\ref{channel_struc}) shows that each path matrix $\mathbf{H}_i$ is a phase-weighted permutation matrix and therefore contains one nonzero entry in each row and column. When the AF-domain locations associated with different paths are distinct, their superposition $\mathbf{H}_{\mathrm{eff}}$ contains $P$ nonzero entries per row and column. By contrast, the multiplication $\mathbf{H}_{\mathrm{eff}}^H\mathbf{H}_{\mathrm{eff}}$ introduces the cross-correlation terms associated with all ordered path pairs. Consequently, each row and column of $\mathbf{G}$ contains at most $2P-1$ nonzero entries. Thus, under integer Doppler conditions, $\mathbf{H}_{\mathrm{eff}}$ generally exhibits a sparser structure than its corresponding Gram matrix.

In practical doubly dispersive channels, the normalized Doppler shifts may contain fractional components. In this case, the energy contributed by each propagation path spreads from its nominal AF-domain position to neighboring entries, and the exact sparsity of $\mathbf{H}_{\mathrm{eff}}$ is no longer preserved. Nevertheless, matched filtering reorganizes the channel representation through the column correlations of $\mathbf{H}_{\mathrm{eff}}$. Specifically, the diagonal elements of $\mathbf{G}$ represent the energies of the corresponding channel columns, whereas its off-diagonal elements characterize the correlations between different columns. Owing to the limited overlap among the dominant AF-domain components, these cross-correlations are typically weaker than the diagonal terms. As a result, although $\mathbf{G}$ may not be strictly sparse under fractional Doppler shifts, its energy is strongly concentrated around the main diagonal. This diagonally dominant and compact structure makes the Ungerboeck observation model particularly suitable for low-complexity IC-based symbol detection.

\begin{figure}
	\centering
	\subfigure[Forney observation model.]{
		\includegraphics[width=0.23\textwidth]{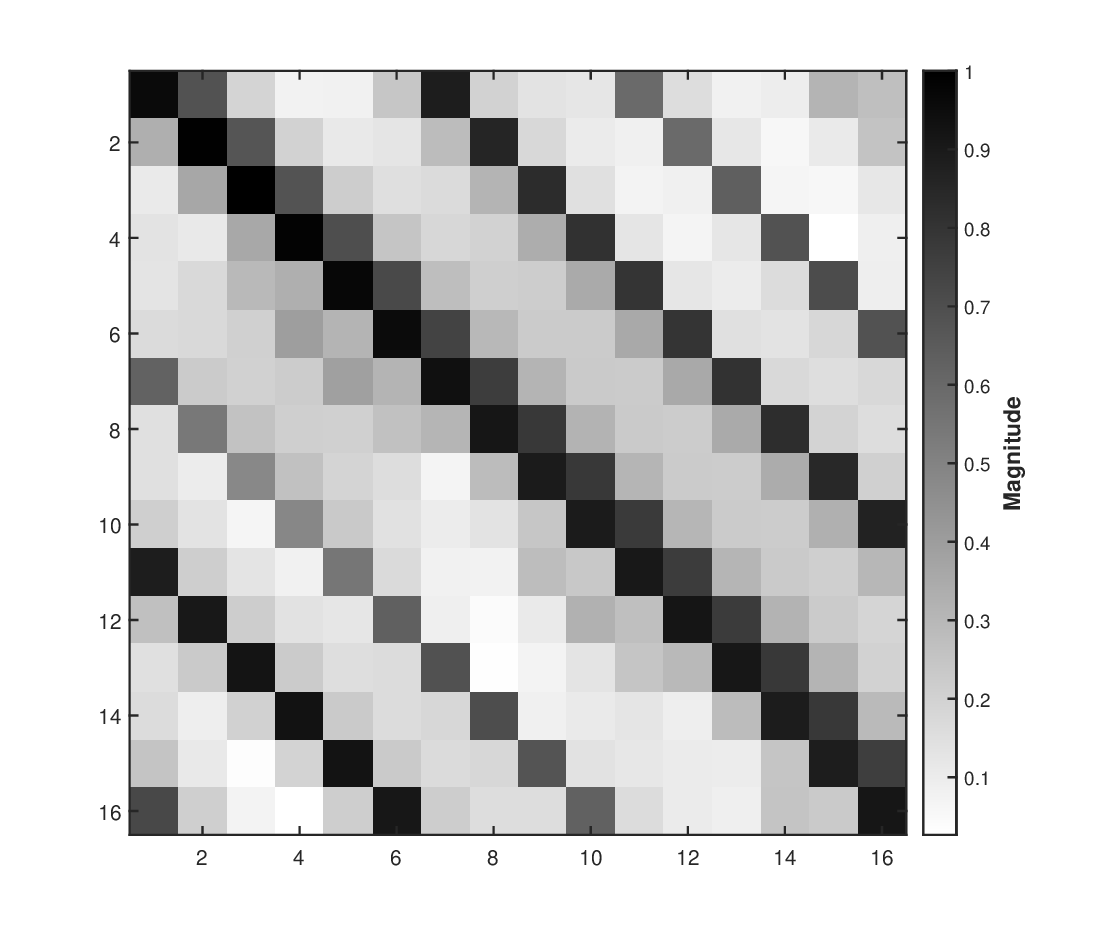}}
	\subfigure[Ungerboeck observation model.]{
		\includegraphics[width=0.23\textwidth]{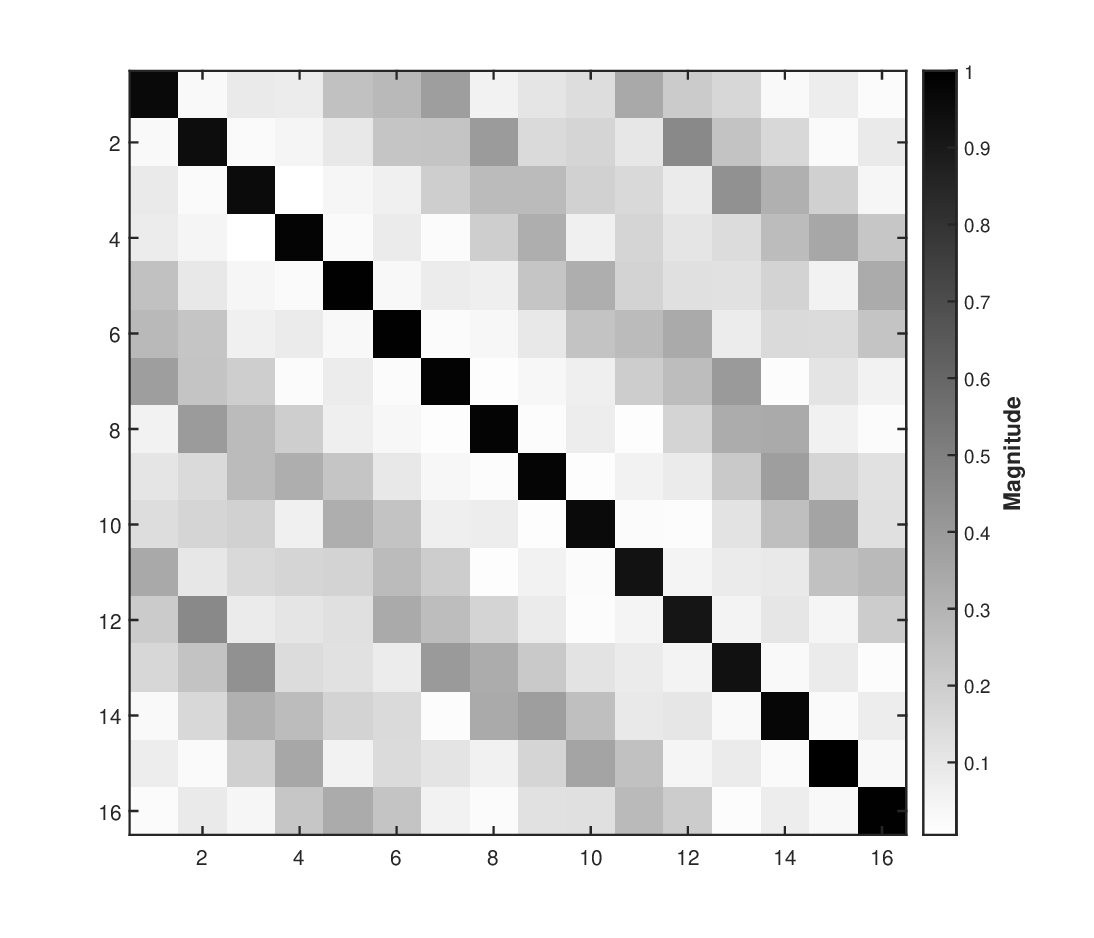}}
	\caption{Comparison of the energy distribution for the channel matrix.}
	\label{fig:1}
\end{figure}
Fig.~\ref{fig:1} compares the energy distributions of the effective Forney channel matrix $\mathbf{H}_{\mathrm{eff}}$ and its corresponding Ungerboeck channel matrix $\mathbf{G}$. Compared with $\mathbf{H}_{\mathrm{eff}}$, $\mathbf{G}$ exhibits substantially stronger energy concentration along its main diagonal, resulting in a more compact channel representation that is well suited for IC-based symbol detection.

\section{Low-Complexity AFDM Symbol Detection based on Interference Cancellation}
\label{Detection Algorithm}

In this section, we develop a low-complexity symbol detector for AFDM under the Ungerboeck observation model. Leveraging the compact representation of $\mathbf{G}$, an IC-based detector is first developed. We then investigate its convergence behavior in high-SNR, revealing the existence of periodic limit cycles, and finally propose an annealing strategy to improve convergence and detection performance.

\subsection{Conventional IC-based Detection Algorithm}

The $m$-th entry of the Ungerboeck observation in (\ref{Ungerboeck model}) can be decomposed into the desired symbol component, inter-symbol interference, and colored noise as
\begin{align}\label{zm}
	z_m
	=
	G_{m,m}x_m
	+
	\xi_m
	+
	\widetilde{w}_m,
\end{align}
where the aggregate interference imposed by the remaining symbols is
\begin{align}\label{xim}
	\xi_m
	=
	\sum_{\substack{n=0,n\neq m}}^{N-1}
	G_{m,n}x_n,
\end{align}
and the colored noise is given by
\begin{align}\label{wm}
	\widetilde{w}_m
	=
	\sum_{n=0}^{N-1}
	H_{\text{eff},m,n}^{*}\overline{w}_n.
\end{align}

As discussed previously, the energy of $\mathbf{G}$ is predominantly concentrated along its main diagonal. Therefore, $G_{m,m}x_m$ generally constitutes the dominant component of $z_m$, whereas the off-diagonal coefficients describe comparatively weaker but potentially non-negligible interference from the other transmitted symbols. This structure enables low-complexity interference cancellation: the interfering symbols are first approximated using their current soft estimates, after which the reconstructed interference is removed from the original observation. Repeating this estimation-and-cancellation procedure progressively suppresses the residual coupling among symbols and improves the reliability of symbol detection~\cite{Conti2026TWC}.

Owing to the strong diagonal energy concentration of $\mathbf{G}$, the residual off-diagonal interference is neglected when evaluating the symbol-wise likelihood. Accordingly, at the $t$-th iteration, the interference-reduced observation $z_m^{[t]}$ is approximated as a complex Gaussian random variable with mean $G_{m,m}x_m$ and variance $\sigma^2_nG_{m,m}$. Its likelihood function can therefore be expressed as
\begin{align}\label{likelihood}
	P^{[t]}(z_m^{[t]}|x_m)
	&\propto
	\mathcal{CN}
	\!\left(
	z_m^{[t]}
	\middle|
	G_{m,m}x_m,
	\sigma^2_nG_{m,m}
	\right)
	\nonumber\\
	&\propto
	\exp
	\!\left(
	-
	\frac{|z_m^{[t]}-G_{m,m}x_m|^2}
	{\sigma^2_nG_{m,m}}
	\right).
\end{align}

Assuming equiprobable constellation symbols, the corresponding soft estimate of $x_m$ is obtained as
\begin{align}\label{posterior}
	\mu_m^{[t]}
	&=
	\mathbb{E}_{P^{[t]}}
	\!\left[
	x_m
	\middle|
	z_m^{[t]}
	\right]
	\notag\\
	&=
	\sum_{a_s\in\mathcal A}
	P^{[t]}(x_m=a_s|z_m^{[t]})a_s.
\end{align}

After computing the soft estimates of all transmitted symbols, the interference affecting the $m$-th observation is reconstructed as
\begin{align}\label{interference}
	\hat{\xi}_m^{[t]}
	=
	\sum_{\substack{n=0,n\neq m}}^{N-1}
	G_{m,n}\mu_n^{[t]}.
\end{align}

The observation for the next iteration is then updated by subtracting the estimated interference from the original matched-filtered observation,The observation used in the subsequent iteration is then generated by subtracting the reconstructed interference from the original matched-filtered observation, i.e,
\begin{align} \label{res}
	z_m^{[t+1]}
	=
	z_m
	-
	\hat{\xi}_m^{[t]},
\end{align}

The soft estimation, interference reconstruction, and observation refinement steps are repeated until the change between two consecutive iterations falls below a prescribed threshold or the maximum number of iterations is reached.
\subsection{Analysis of the Conventional IC-based Detector}
Although the IC-based detector introduced in~\cite{Conti2026TWC} offers low computational complexity, its detection performance may deteriorate in the high-SNR regime. To investigate the underlying mechanism, we first write the normalized posterior probability of $x_m=a_s$ at the $t$-th iteration as
\begin{align}\label{spelikelihood}
	P^{[t]}(x_m=a_s|z_m^{[t]})
	=
	\frac{\exp\!\left(
		-\frac{|z_m^{[t]}-G_{m,m}a_s|^2}
		{\sigma^2_nG_{m,m}}
		\right)}
	{\sum\limits_{a_s\in\mathcal A}
		\exp\!\left(
		-\frac{|z_m^{[t]}-G_{m,m}a_s|^2}
		{\sigma^2_nG_{m,m}}
		\right)}.
\end{align}

To characterize the asymptotic behavior, consider a set of real-valued metrics $\{x_1,x_2,\ldots,x_n\}$ with a unique minimum attained at $x_k$, i.e.,
$
x_k<x_i, \forall i\neq k.
$
Then, the following limit holds
\begin{align}\label{limit}
	\lim_{a\rightarrow0^{+}}
	\frac{\exp\!\left(-\frac{x_i}{a}\right)}
	{\sum\limits_{j=1}^{n}
		\exp\!\left(-\frac{x_j}{a}\right)}
	=
	\begin{cases}
		1,& i=k,\\
		0,& i\neq k.
	\end{cases}
\end{align}
This result indicates that, as the scaling factor $a$ approaches zero, the exponential term corresponding to the minimum metric increasingly dominates the denominator, whereas the relative contributions of all other terms vanish asymptotically. Consequently, the posterior distribution converges to a one-hot distribution concentrated on the minimum-metric candidate.

Therefore, in the high-SNR regime, i.e., as $\sigma_n^2\rightarrow0^{+}$, the posterior distribution in (\ref{spelikelihood}) becomes increasingly concentrated on the constellation point that minimizes the Euclidean-distance metric. Consequently, the posterior distribution converges to a one-hot distribution, and the soft estimate in (\ref{posterior}) asymptotically reduces to the following hard decision
\begin{align}\label{hard estimate}
	\mu_m^{[t]}
	&=
	\arg\min_{a_s\in\mathcal A}
	|z_m^{[t]}-G_{m,m}a_s|^2 \\ \notag
	&\triangleq\arg\min_{a_s\in\mathcal A}
	|\widetilde{z}_m^{[t]}-a_s|^2, 
\end{align}
where $\widetilde{z}_m^{[t]} = {z}_m^{[t]} / G_{m,m}$.

To gain further insight into the performance degradation of the conventional IC detector in the high-SNR regime, we analyze the evolution of the estimation error over successive iterations. Specifically, the estimation error at the $t$-th iteration is defined as
\begin{align}\label{error definition}
	\mathbf{e}^{[t]} = \mathbf{x} - \boldsymbol{\mu}^{[t]}.
\end{align}
We further decompose the channel matrix $\mathbf{G}$ into its diagonal and off-diagonal components
\begin{align}\label{comp}
	\mathbf{G} = \mathbf{D} + \mathbf{S},
\end{align}
where $\mathbf{D} = \operatorname{diag}\{\mathbf{G}\}$ contains the diagonal entries of $\mathbf{G}$, and $\mathbf{S} = \mathbf{G} - \mathbf{D}$ contains its off-diagonal interference terms.

To facilitate the analysis, the additive noise is neglected, which is a reasonable approximation in the high-SNR regime. According to the update rule in (\ref{res}), the observation used at the $(t+1)$-th iteration can be written as
\begin{align}
	\mathbf{z}^{[t+1]} = \mathbf{z}-\mathbf{S}\boldsymbol{\mu}^{[t]}= \mathbf{D}\mathbf{x} + \mathbf{S}\mathbf{e}^{[t]}.
\end{align}
Normalizing both sides by $\mathbf{D}$ yields
\begin{align}\label{next observation}
	\widetilde{\mathbf{z}}^{[t+1]} = \mathbf{x} + \mathbf{\Omega}\mathbf{e}^{[t]},
\end{align}
where $\mathbf{\Omega} = \mathbf{D}^{-1}\mathbf{S}$.

Based on (\ref{hard estimate}), the symbol estimates at the $(t+1)$-th iteration are obtained by applying element-wise hard-decision slicing
\begin{align} \label{next mu}
	\boldsymbol{\mu}^{[t+1]} = \mathcal{T}_{\mathcal{A}}\left(\widetilde{\mathbf{z}}^{[t+1]}\right),
\end{align}
where $\mathcal{T}_{\mathcal A}(\cdot)$ denotes the element-wise nearest-neighbor projection onto the constellation alphabet $\mathcal A$.

Substituting (\ref{next observation}) into (\ref{next mu}) and using the error definition in (\ref{error definition}), the evolution of the estimation error can be compactly expressed as
\begin{align} \label{iterative equation}
	\mathbf{e}^{[t+1]} = \mathbf{x}-\mathcal{T}_{\mathcal{A}}\left(\mathbf{x}+\mathbf{\Omega}\mathbf{e}^{[t]}\right).
\end{align}
This recursion defines a nonlinear discrete-time dynamical system governing the error evolution. In particular, the error at each iteration is determined solely by the preceding error through the normalized interference matrix $\boldsymbol{\Omega}$ and the nonlinear hard-decision operator. Therefore, the convergence behavior of the conventional IC detector can be characterized by examining the stability properties of this system.

Since the hard-decision operator $\mathcal{T}_{\mathcal A}(\cdot)$ is applied element-wise in (\ref{iterative equation}), the resulting symbol estimate belongs to $\mathcal A^{N \times 1}$. Moreover, since the transmitted symbol vector satisfies $\mathbf{x}\in\mathcal A^{N \times 1}$, each component of the estimation error belongs to the finite difference set $\mathcal{E}=\{a_i-a_j\,|\,a_i,a_j\in\mathcal{A}\}$.
Consequently, the error vector evolves within the finite state space $	\mathbf e^{[t]}
\in
\mathcal E^N$,
whose cardinality satisfies $|\mathcal E^N|
=
|\mathcal E|^N
<
\infty$.

According to the pigeonhole principle, among any $|\mathcal E|^N+1$ consecutive error states, there must exist two indices $t_1<t_2$ such that $\mathbf e^{[t_1]}
=
\mathbf e^{[t_2]}$. Because (\ref{iterative equation}) defines a deterministic mapping, equality of these two states implies that all subsequent states repeat identically.
Let $L=t_2-t_1$, we can obtain
\begin{align} \label{Lperiod}
	\mathbf e^{[t+L]}
	=
	\mathbf e^{[t]},
	\quad
	t\ge t_1.
\end{align}
Therefore, the estimation-error sequence generated by (\ref{iterative equation}) eventually enters a periodic orbit. When $L=1$, this orbit reduces to a fixed point, indicating that the detector has converged, although the fixed point need not correspond to the correct symbol vector. By contrast, when $L>1$, the detector becomes trapped in a finite-period limit cycle and fails to converge to a stationary estimate.

\begin{figure}
	\centering
	\includegraphics[width=0.5\textwidth]{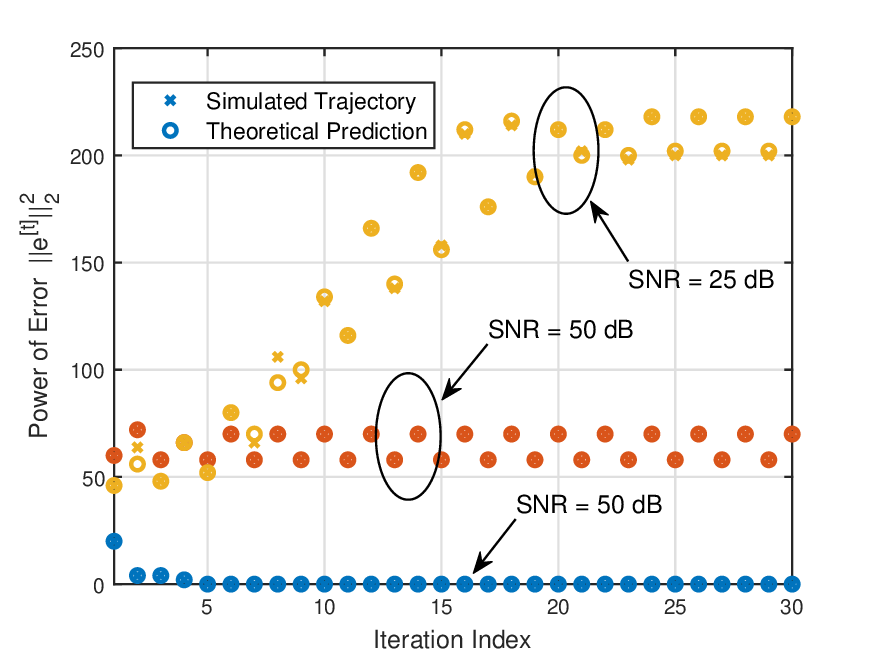}
	\caption{The error evolution of conventional IC-based detection algorithm.}\label{fig:2}
\end{figure}

The estimation-error evolution of the conventional IC detector is evaluated in Fig.~\ref{fig:2}. At an SNR of 50 dB, where noise is negligible, the theoretical prediction obtained from (\ref{iterative equation}) closely matches the simulation results, validating the proposed error-evolution model. The blue curve converges to the zero-error fixed point, whereas the orange curve enters a finite-period limit cycle, confirming the two asymptotic behaviors predicted by (\ref{Lperiod}). At 25 dB, the theoretical and simulated results remain in close agreement, with minor deviations caused by the non-negligible noise.


\subsection{The IC-based Detector with Annealing Schedule}

As analyzed above, in the high-SNR regime, the posterior distribution becomes excessively sharp, causing the IC detector to behave as a hard-decision dynamical system and potentially become trapped in a finite-period limit cycle. To alleviate this problem, we introduce a temperature parameter to smooth the posterior distribution and enhance the detector’s ability to escape undesirable periodic orbits. However, maintaining an excessively high temperature may overflatten the posterior distribution and lead to persistent random-walk behavior~\cite{Hassibi2014TSP}. Therefore, a gradual annealing strategy is adopted to balance stochastic exploration and reliable convergence. 

Specifically, a temperature parameter $T_t$ is introduced into the posterior probability, yielding
\begin{align}\label{Tlikelihood}
	P^{[t]}(x_m=a_s|z_m^{[t]})
	=
	\frac{\exp\!\left(
		-\frac{|z_m^{[t]}-G_{m,m}a_s|^2}
		{T_t\sigma^2_nG_{m,m}}
		\right)}
	{\sum\limits_{a_s\in\mathcal A}
		\exp\!\left(
		-\frac{|z_m^{[t]}-G_{m,m}a_s|^2}
		{T_t\sigma^2_nG_{m,m}}
		\right)},
\end{align}
where $T_t=\text{max}\{T_0\beta^{t-1},1\}$ is the temperature at the $t$-th iteration, $T_0$ is the initial temperature, and $0<\beta<1$ denotes the cooling factor.

Furthermore, to enhance convergence stability and suppress oscillations during the iterative process, a damping mechanism is incorporated into the observation update~\cite{Conti2026TWC}
\begin{align} \label{res2}
	z_m^{[t+1]}
	=
	\eta\left(z_m
	-
	\hat{\xi}_m^{[t]}\right)
	+
	(1-\eta)z_m^{[t]},
\end{align}
where $0<\eta\le1$ denotes the damping factor.

The overall IC-based Detector with annealing schedule is summarized in Algorithm~\ref{alg:IC}.
\renewcommand{\algorithmicrequire}{\textbf{Input:}}
\renewcommand{\algorithmicensure}{\textbf{Output:}}
\begin{algorithm}[!htbp]
	\caption{The Annealed IC Symbol Detector}
	\label{alg:IC}
	\begin{algorithmic}[1]
		\Require Received symbols $\mathbf{z}$, channel matrix $\mathbf{G}$, noise power $\sigma_n^2$, initial temperature $T_0$, cooling rate $\beta$, damping factor $\eta$, convergence threshold $\epsilon$, and maximum number of iterations $N_{max}$.
		\State \textbf{Initialization:} $T_t=T_0$.
		\For{$t = 0$ to $N_{max}-1$}
		\For{$m = 0$ to $N-1$}
		\State Compute the symbol posterior distribution $P^{[t]}(x_m=a_s|z_m^{[t]})$ based on (\ref{Tlikelihood});
		\State Compute the soft symbol estimate $\mu_m^{[t]}$ based on (\ref{posterior});
		\State Compute the estimate of the interference term $\hat{\xi}_m^{[t]}$ based on (\ref{interference});
		\State Update $z_m^{[t+1]}$ based on (\ref{res2});
		\EndFor
		\If{$\Vert\mathbf{z}^{[t+1]}-\mathbf{z}^{[t]}\Vert_2^2<\epsilon$}
		\State{Break; }
		\EndIf
		\State Update $T_t=\text{max}\{T_0\beta^{t-1},1\}$;
		\EndFor
		\Ensure $P^{[t]}(x_m=a_s|z_m^{[t]}), \forall m = 0, \dots, N-1$.
	\end{algorithmic}
\end{algorithm}

\section{Annealing Parameter Design for IC-based Detector}
\label{sec:dynamics}
In this section, we formulate the annealed IC detector as a temperature-dependent nonlinear dynamical system and analyze its fixed points. Based on this analysis, we characterize the influence of temperature on the fixed-point evolution of the detector. We then derive principled strategies for selecting the cooling factor and initial temperature, thereby balancing convergence stability, tracking capability, and computational efficiency.

\subsection{Dynamical System Formulation of the Annealed IC-based Detector}
\label{subsec:equivalent_state}

As shown in Algorithm~\ref{alg:IC}, the update rule of the annealed IC detector at the $t$-th iteration can be expressed in matrix form as
\begin{align}
	\mathbf{z}^{[t+1]}=\left(1-\alpha\right)\mathbf{z}^{[t]}+\alpha\left[\mathbf{z}-\mathbf{S}\mathbf{F}_{T_t}\left(\mathbf{z}^{[t]}\right)\right],
\end{align}
where $\mathbf{F}_{T_t}(\cdot)$ denotes the posterior-mean mapping defined in (\ref{posterior}) at temperature $T_t$.

To characterize the fixed-point structure of the detector, we consider its dynamics at a prescribed temperature $T_t$. If the iteration converges to a fixed point $\mathbf{z}^{\star}$, then
\begin{align}
	\mathbf{z}^{\star}=\left(1-\alpha\right)\mathbf{z}^{\star}+\alpha\left[\mathbf{z}-\mathbf{S}\mathbf{F}_{T_t}\left(\mathbf{z}^{\star}\right)\right].
\end{align}
For $0<\alpha\leq1$, the fixed-point condition can be simplified as
\begin{align} \label{fixed point 0}
	\mathbf{z}^{\star}=\mathbf{z}-\mathbf{S}\mathbf{F}_{T_t}\left(\mathbf{z}^{\star}\right).
\end{align}
Notably, the damping factor $\alpha$ does not appear in (\ref{fixed point 0}). Therefore, at a given temperature, damping affects the convergence trajectory, and stability of the iterative process, but does not alter the fixed-point set.

The following analysis focuses on unit-energy QPSK with
$\mathcal{A}=\{(\pm1\pm j)/\sqrt{2}\}$; nevertheless, the analytical framework can be extended to other constellation alphabets. For QPSK, the $k$-th component of the posterior-mean mapping is given by
\begin{align}
	F_{T_t,k}\left(x\right)=\frac{1}{\sqrt{2}}\operatorname{tanh}\left(\frac{\operatorname{Re}\left\{x\right\}}{\sqrt{2}\sigma^2_nT_t}\right)+\frac{j}{\sqrt{2}}\operatorname{tanh}\left(\frac{\operatorname{Im}\left\{x\right\}}{\sqrt{2}\sigma^2_nT_t}\right).
\end{align}

Defining $\mathbf{m}^{\star}=\mathbf{F}_{T_t}\left(\mathbf{z}^{\star}\right)$, and substituting (\ref{fixed point 0}) into this definition yield
\begin{align} \label{fixed point 1}
	\mathbf{m}^{\star}=\mathbf{F}_{T_t}\left(\mathbf{z}-\mathbf{S}\mathbf{m}^{\star}\right),
\end{align}
which serves as the fixed-point equation for the subsequent analysis.

To facilitate the derivation, we introduce the real-valued representations of an arbitrary complex vector $\mathbf{a}$ and complex matrix $\mathbf{A}$ as
\begin{align}
	\mathcal{R}(\mathbf{a})\triangleq
	\begin{bmatrix}{\rm Re}\{\mathbf{a}\}\\{\rm Im}\{\mathbf{a}\}\end{bmatrix},
	\quad
	\mathcal{R}(\mathbf{A})\triangleq
	\begin{bmatrix}
		{\rm Re}\{\mathbf{A}\}&-{\rm Im}\{\mathbf{A}\}\\
		{\rm Im}\{\mathbf{A}\}& {\rm Re}\{\mathbf{A}\}
	\end{bmatrix}.
	\label{eq:realification}
\end{align}
Let $\mathbf{r}_0=\mathcal{R}(\mathbf{z})$,
$\mathbf{m}_{\rm R}^{\star}=\mathcal{R}(\mathbf{m}^{\star})$, and $\mathbf{S}_{\rm R}=\mathcal{R}(\mathbf{S})$.  Since $\mathbf{S}$ is Hermitian, we have $\lambda_{\min}\left(\mathbf{S}\right)=\lambda_{\min}\left(\mathbf{S}_{\rm R}\right)$.

Using the above real-valued representations, the fixed-point equation in (\ref{fixed point 1}) can be rewritten as
\begin{align} \label{fixed point 2}
	\mathbf{m}_{\rm R}^\star=\frac{1}{\sqrt{2}} \operatorname{tanh}\left[\frac{\mathbf{r}_0-\mathbf{S}_{\rm R} \mathbf{m}_{\rm R}^\star}{\sqrt{2} \sigma_n^2 T_t}\right],
\end{align}
where $\operatorname{tanh}(\cdot)$ is applied element-wise.

Applying the element-wise inverse hyperbolic tangent to both sides of (\ref{fixed point 2}) yields
\begin{align} \label{fixed point 3}
\mathbf{S}_{\rm R} \mathbf{m}_{\rm R}^\star+\sqrt{2} \sigma_n^2 T_t\operatorname{artanh}\left(\sqrt{2}\mathbf{m}_{\rm R}^\star\right)-\mathbf{r}_0=\mathbf{0},
\end{align}
which can be interpreted as the stationarity condition of a temperature-dependent potential function $V_{T_t}(\mathbf{m}_{\rm R})$, whose gradient is given by
\begin{align} \label{potential function}
	\nabla V_{T_t}\left(\mathbf{m}_{\rm R}\right)=\mathbf{S}_{\rm R} \mathbf{m}_{\rm R}+\sqrt{2} \sigma_n^2 T_t\operatorname{artanh}\left(\sqrt{2}\mathbf{m}_{\rm R}\right)-\mathbf{r}_0.
\end{align}
Accordingly, the Hessian matrix of $V_{T_t}(\mathbf{m}_{\rm R})$ is
\begin{align}
	\mathbf{H}_T(\mathbf{m}_{\rm R})
	=\nabla^2V_T(\mathbf{m}_{\rm R})
	=\mathbf{S}_{\rm R}+2\sigma^2_nT_t
	\operatorname{diag}\!\left(\frac{1}{1-2m_{\text{R},i}^2}\right).
	\label{eq:hessian}
\end{align}
Under the QPSK posterior-mean mapping, each component satisfies
$|m_{{\rm R},i}|<1/\sqrt{2}$, and hence
$(1-2m_{\text{R},i}^2)^{-1}\ge 1$. The Hessian therefore satisfies
\begin{align}
	\mathbf{H}_T(\mathbf{m}_{\rm R})\succeq
	\mathbf{S}_{\rm R}+2\sigma_n^2T_t\mathbf{I}.
	\label{eq:hessian_bound}
\end{align}
Since $\lambda_{\min}(\mathbf{S}_{\rm R})
=\lambda_{\min}(\mathbf{S})$, a lower bound of the minimum eigenvalue of the Hessian can be expressed as
\begin{align}
	C_t = \lambda_{\min}(\mathbf{S})+2\sigma^2_nT_t
	\label{eq:strict_convex_condition}.
\end{align}
This lower bound provides a theoretical basis for selecting the initial temperature $T_0$, as discussed in the subsection \ref{subsec:jacobian}.

\subsection{Fixed-Point Tracking under Annealing}
\label{subsec:fixed_point_motion}

As the temperature decreases during annealing, the corresponding fixed point varies across iterations. Therefore, the detector must not only converge toward a stable fixed point but also track its temperature-dependent movement. In this subsection, we analyze the displacement between consecutive fixed points and characterize its dependence on the cooling factor $\beta$. Let $\mathbf{m}_{\rm R}^{\star}$ denote a fixed-point branch satisfying $\nabla V_T\left(\mathbf{m}_{\rm R}^{\star}\right) = \mathbf{0}$. Assume that $\mathbf{m}^{\star}_{\rm R}$ is differentiable with respect to $T$ and that
$\mathbf{H}_T^\star\triangleq\mathbf{H}_T(\mathbf{m}^\star_{\rm R})$ is nonsingular.
Differentiating the stationarity condition (\ref{potential function}) with respect to $T$ gives
\begin{align}
	\mathbf{H}_T^{\star}
	\frac{d\mathbf{m}^{\star}_{\rm R}}{dT}
	+
	\sqrt{2}\sigma_n^2
	\operatorname{artanh}\left(
	\sqrt{2}\mathbf{m}^{\star}_{\rm R}
	\right)=\mathbf{0}.
\end{align}
Hence, the temperature sensitivity of the fixed point is given by
\begin{align}
	\frac{d\mathbf{m}^\star}{dT}
	=-\sqrt{2}\sigma_n^2\left(\mathbf{H}_T^{\star}\right)^{-1}
	\operatorname{artanh}\left(
	\sqrt{2}\mathbf{m}^{\star}_{\rm R}
	\right).
	\label{eq:fixed_point_derivative}
\end{align}

During the active cooling stage, the temperature is updated according to $T_{t+1}=\beta T_t$. We define the displacement between two consecutive temperature-dependent fixed points as
\begin{align}
	\bm{\delta}_t\triangleq
	\mathbf{m}^\star_{\rm R}(T_{t+1})-\mathbf{m}^\star_{\rm R}(T_t).
	\label{eq:fixed_point_increment_definition}
\end{align}
Applying a first-order Taylor expansion around $T_t$, we obtain
\begin{align}
	\bm{\delta}_t
	=\;-(1-\beta)T_t\left.\frac{d\mathbf{m}_{\rm R}^\star}{dT}\right|_{T=T_t} +\mathcal{O}\!\left((1-\beta)^2T_t^2\right).
	\label{eq:fixed_point_increment}
\end{align}

Accordingly, the magnitude of the fixed-point displacement approximately satisfies
\begin{align}\label{eq:fixed_point_increment_bound}
	\left\Vert\boldsymbol{\delta}_t\right\Vert
	\lesssim
	\sqrt{2}(1-\beta)T_t\sigma_n^2
	\left\Vert\left(\mathbf{H}_{T_t}^{\star}\right)^{-1}\right\Vert
	\left\Vert
	\operatorname{artanh}\left(
	\sqrt{2}\mathbf{m}^{\star}_{\rm R}(T_t)
	\right)
	\right\Vert,
\end{align}
which shows that the displacement between consecutive fixed points scales approximately linearly with $1-\beta$. 

Therefore, selecting $\beta$ sufficiently close to one reduces the movement of the target fixed point and allows the detector trajectory to track it more accurately. 
Conversely, an excessively small $\beta$ causes the target fixed point to move too rapidly, such that the detector may fail to remain within its attraction region. However, choosing $\beta$ overly close to one leads to slow cooling and increases the required number of iterations. The cooling-factor design should therefore balance fixed-point tracking accuracy against convergence speed.


\subsection{Local Tracking Behavior and Time-Varying Jacobian Analysis}
\label{subsec:jacobian}

We next investigate the local tracking behavior of the detector
through a time-varying Jacobian analysis.
The resulting tracking-error recursion characterizes the combined effects
of local stability and fixed-point motion, thereby providing a theoretical
basis for selecting the initial temperature $T_0$.

For the local fixed-point analysis, the detector dynamics can be represented as
\begin{align}
	\mathbf{m}_{\rm R}^{[t+1]}
	&=
	\boldsymbol{\Psi}_{T_t}
	\left(\mathbf{m}_{\rm R}^{[t]}\right)
	\notag\\
	&\triangleq
	(1-\alpha)\mathbf{m}_{\rm R}^{[t]}
	+
	\frac{\alpha}{\sqrt{2}}
	\operatorname{tanh}\left[
	\frac{
		\mathbf{r}_0-\mathbf{S}_{\rm R}\mathbf{m}_{\rm R}^{[t]}
	}{
		\sqrt{2}\sigma_n^2T_t
	}
	\right],
	\label{eq:m_mapping}
\end{align}
At temperature $T_t$, the fixed point satisfies
$
	\boldsymbol{\Psi}_{T_t}
	\left(\mathbf{m}_{\rm R}^{\star}(T_t)\right)
	=
	\mathbf{m}_{\rm R}^{\star}(T_t)
$.
Define the tracking error relative to the fixed point as
\begin{align}
	\mathbf{e}_t
	\triangleq
	\mathbf{m}_{\rm R}^{[t]}
	-
	\mathbf{m}_{\rm R}^{\star}(T_t).
	\label{eq:tracking_error_0}
\end{align}
The tracking error at the next iteration can then be expressed as
\begin{align}
	\mathbf{e}_{t+1}
	=
	\boldsymbol{\Psi}_{T_t}
	\left(
	\mathbf{m}_{\rm R}^{\star}(T_t)
	+
	\mathbf{e}_{t}
	\right)
	-
	\mathbf{m}_{\rm R}^{\star}(T_{t+1}).
	\label{eq:tracking_error_1}
\end{align}

Assuming that $\|\mathbf{e}_{t}\|_2$ is sufficiently small,
a first-order Taylor expansion around
$\mathbf{m}_{\rm R}^{\star}(T_t)$ gives
\begin{align}
	\boldsymbol{\Psi}_{T_t}
	\left(
	\mathbf{m}_{\rm R}^{\star}(T_t)
	+
	\mathbf{e}_t
	\right)
	&=
	\boldsymbol{\Psi}_{T_t}
	\left(
	\mathbf{m}_{\rm R}^{\star}(T_t)
	\right)
	+
	\mathbf{J}_t\mathbf{e}_t+
	\mathcal{O}
	\left(
	\|\mathbf{e}_t\|_2^2
	\right)
	\notag\\
	&=
	\mathbf{m}_{\rm R}^{\star}(T_t)
	+
	\mathbf{J}_t\mathbf{e}_t
	+
	\mathcal{O}
	\left(
	\|\mathbf{e}_t\|_2^2
	\right),
	\label{eq:tracking_error_2}
\end{align}
where
\begin{align}
	\mathbf{J}_t
	&\triangleq
	\left.
	\frac{
		\partial\boldsymbol{\Psi}_{T_t}(\mathbf{m}_{\rm R})
	}{
		\partial\mathbf{m}_{\rm R}
	}
	\right|_{
		\mathbf{m}_{\rm R}
		=
		\mathbf{m}_{\rm R}^{\star}(T_t)
	}
	\notag\\
	&=
	(1-\alpha)\mathbf{I}
	-
	\alpha\mathbf{D}_t\mathbf{S}_{\rm R}
	\label{eq:jacobian}
\end{align}
is the time-varying Jacobian matrix evaluated at the instantaneous fixed
point. The diagonal matrix $\mathbf{D}_t$ is given by
\begin{align}
	\mathbf{D}_t
	=
	\operatorname{diag}
	\left[
	\frac{1}{2\sigma_n^2T_t}
	\operatorname{sech}^{2}
	\left(
	\frac{r_{t,i}^{\star}}
	{\sqrt{2}\sigma_n^2T_t}
	\right)
	\right],
	\label{eq:D_t}
\end{align}
where
\begin{align}
	\mathbf{r}_t^{\star}
	\triangleq
	\mathbf{r}_0
	-
	\mathbf{S}_{\rm R}
	\mathbf{m}_{\rm R}^{\star}(T_t).
	\label{eq:r_fixed_point}
\end{align}

Substituting (\ref{eq:tracking_error_2}) into
(\ref{eq:tracking_error_1}) yields the local tracking-error recursion
\begin{align}
	\mathbf{e}_{t+1}
	=
	\mathbf{J}_t\mathbf{e}_t
	-
	\boldsymbol{\delta}_t
	+
	\mathcal{O}
	\left(
	\|\mathbf{e}_t\|_2^2
	\right),
	\label{eq:tracking_error_recursion}
\end{align}
which shows that the tracking-error
evolution is jointly governed by the local Jacobian $\mathbf{J}_t$ and
the fixed-point displacement $\boldsymbol{\delta}_t$. Specifically,
$\rho(\mathbf{J}_t)<1$ ensures local contraction at temperature $T_t$,
whereas a sufficiently small $\|\boldsymbol{\delta}_t\|_2$ allows the
detector to accurately track the moving fixed point. Conversely, if
$\rho(\mathbf{J}_t)\geq1$, local perturbations may be amplified, leading
to tracking failure.

\begin{figure*}[!htbp]
	\centering
	\subfigure[Evolution of the eigenvalue lower bound $C_t$.]{
		\label{fig:convexity_margin}
		\includegraphics[width=0.32\textwidth]{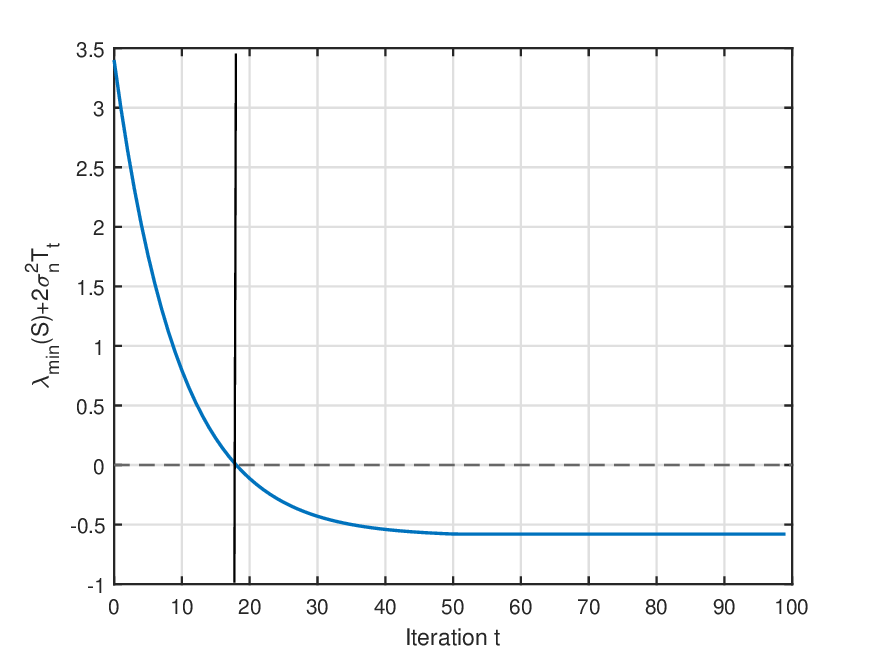}}
	\hfill
	\subfigure[Evolution of the Jacobian spectral radius
	$\rho(\mathbf{J}_t)$.]{
		\label{fig:jacobian_radius}
		\includegraphics[width=0.32\textwidth]{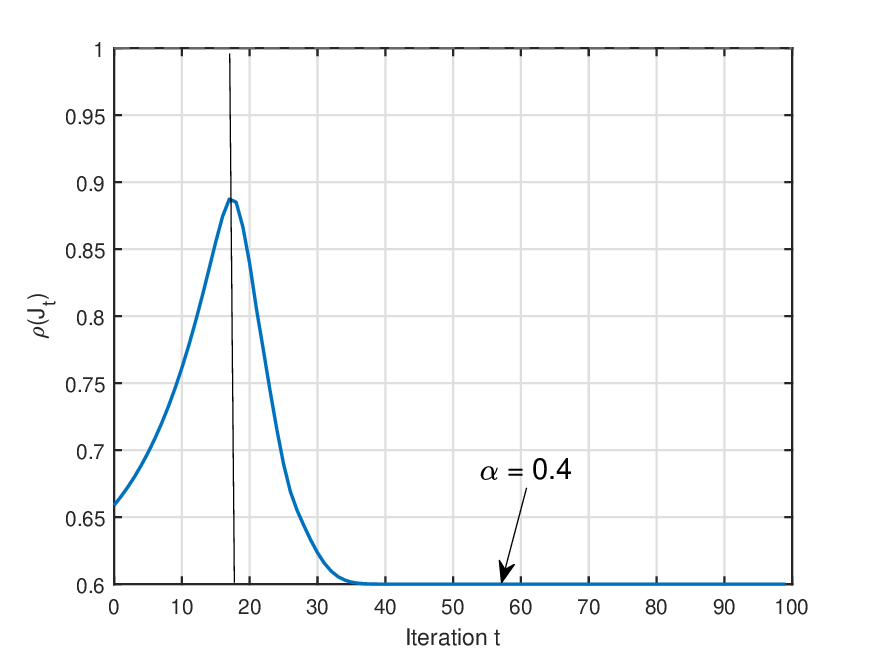}}
	\hfill
	\subfigure[Actual and first-order-predicted fixed-point displacement
	$\|\boldsymbol{\delta}_t\|_2$.]{
		\label{fig:fixed_point_displacement}
		\includegraphics[width=0.32\textwidth]{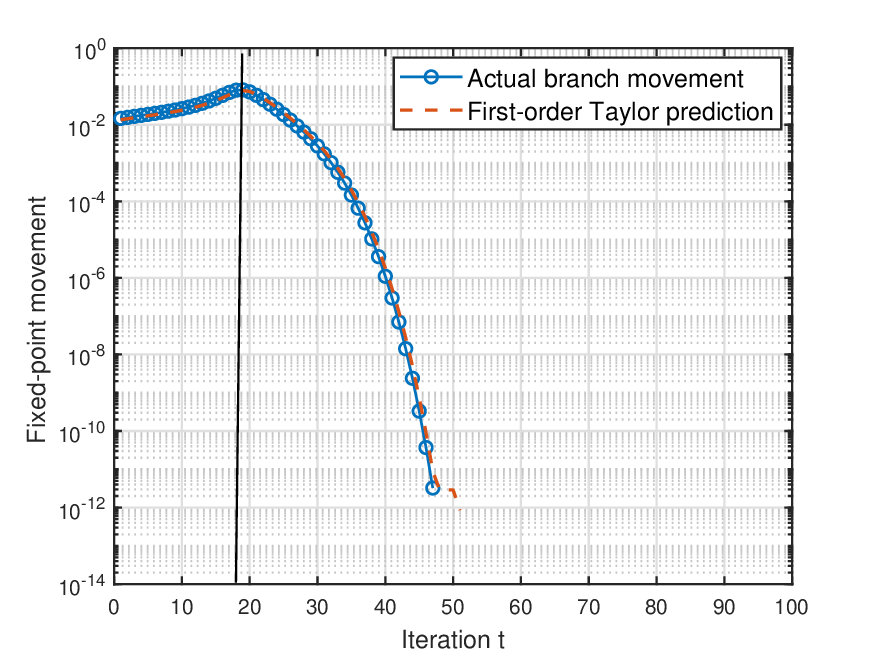}}
	\caption{The temperature-dependent fixed-point and
		Jacobian analyses during the annealing process with $\beta=0.9$.}
	\label{fig:temperature_dynamics}
\end{figure*}

To guide the selection of the initial temperature, Fig.~\ref{fig:temperature_dynamics}, which is shown at the top of the next page, illustrates the evolution of the $C_t$, the Jacobian spectral radius $\rho(\mathbf{J}_t)$, and the displacement $\|\bm{\delta}_t\|_2$ between consecutive fixed points during annealing. It can be observed that the zero crossing of $C_t$ occurs in close proximity to the turning points of $\rho(\mathbf{J}_t)$ and $\|\bm{\delta}_t\|_2$, suggesting that all three characterize the same transition in the temperature-dependent dynamics.

To explain the close proximity of these characteristic points, we first
introduce an auxiliary variable evaluated at the instantaneous fixed point
\begin{align}
	u_{t,i}^{\star}
	\triangleq
	\frac{r_{t,i}^{\star}}
	{\sqrt{2}\sigma_n^2T_t}.
	\label{eq:auxiliary_u}
\end{align}
Accordingly, the diagonal matrix in (\ref{eq:D_t}) can be rewritten as
\begin{align}
	\mathbf{D}_t
	=
	\operatorname{diag}\left[
	\frac{
		\operatorname{sech}^2\left(u_{t,i}^{\star}\right)
	}{
		2\sigma_n^2T_t
	}
	\right].
	\label{eq:D_m_relation}
\end{align}

The annealing trajectory in Fig.~\ref{fig:temperature_dynamics} can then
be divided into two regions according to the sign of $C_t$.

\emph{1) High-temperature region with $C_t>0$:}
In this region, the
posterior distribution remains relatively smooth, i.e,
$
	\operatorname{sech}^2\left(u_{t,i}^{\star}\right)
	\approx 1
$. We can obtain
\begin{align}
	\mathbf{D}_t
	\approx
	\frac{1}{2\sigma_n^2T_t}\mathbf{I}.
	\label{eq:high_temperature_approximation}
\end{align}
Substituting (\ref{eq:high_temperature_approximation}) into
(\ref{eq:jacobian}) gives
\begin{align}
	\mathbf{J}_t
	\approx
	(1-\alpha)\mathbf{I}
	-
	\frac{\alpha}{2\sigma_n^2T_t}\mathbf{S}_{\rm R}.
\end{align}

When the mode associated with $\lambda_{\min}(\mathbf{S})<0$ dominates,
we can obtain
\begin{align}
	\rho(\mathbf{J}_t)
	\approx
	1-\alpha
	-
	\frac{
		\alpha\lambda_{\min}(\mathbf{S})
	}{
		2\sigma_n^2T_t
	}.
	\label{eq:critical_jacobian_eigenvalue}
\end{align}
As $T_t$ decreases, the contribution of the negative-eigenvalue mode
increases. In particular, when
$
	2\sigma_n^2T_t
	\approx
	-\lambda_{\min}(\mathbf{S})
$,
the right-hand side of (\ref{eq:critical_jacobian_eigenvalue}) approaches
unity. This condition is equivalent to $C_t\approx0$. 

\begin{figure*}[!htbp]
	\centering
	\subfigure[Evolution of the eigenvalue lower bound $C_t$.]{
		\label{fig:convexity_margin2}
		\includegraphics[width=0.32\textwidth]{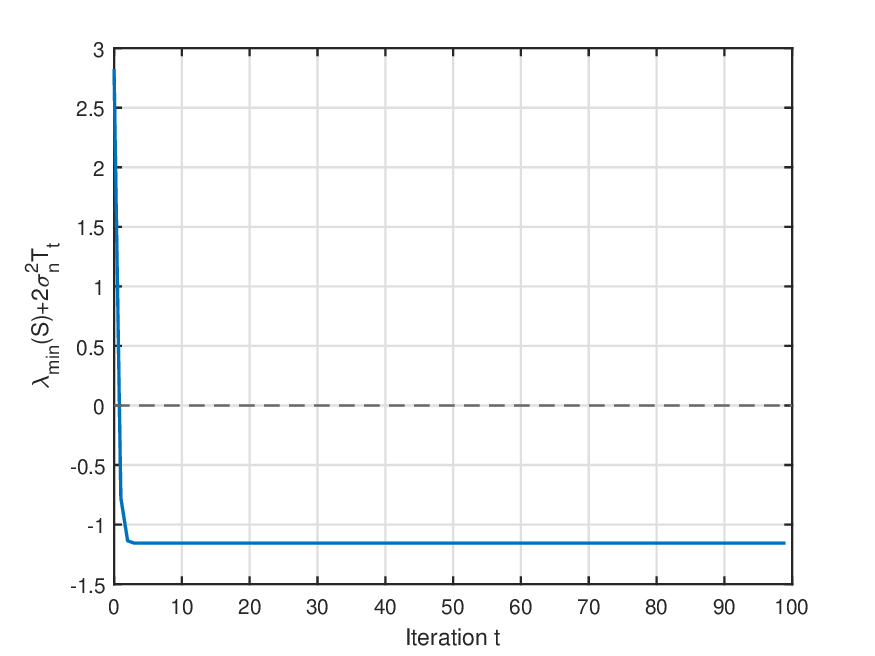}}
	\hfill
	\subfigure[Evolution of the Jacobian spectral radius
	$\rho(\mathbf{J}_t)$.]{
		\label{fig:jacobian_radius2}
		\includegraphics[width=0.32\textwidth]{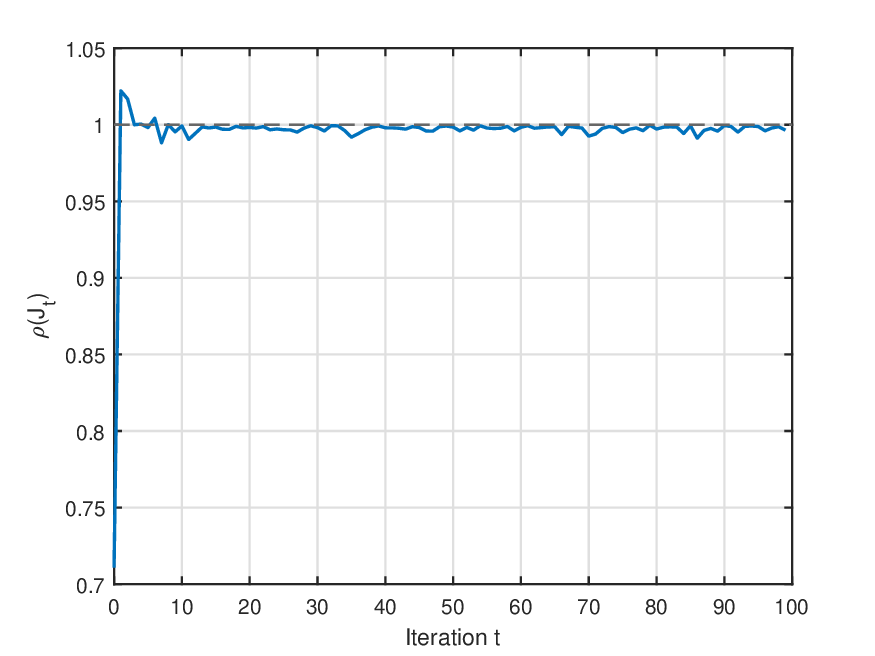}}
	\hfill
	\subfigure[Actual and first-order-predicted fixed-point displacement
	$\|\boldsymbol{\delta}_t\|_2$.]{
		\label{fig:fixed_point_displacement2}
		\includegraphics[width=0.32\textwidth]{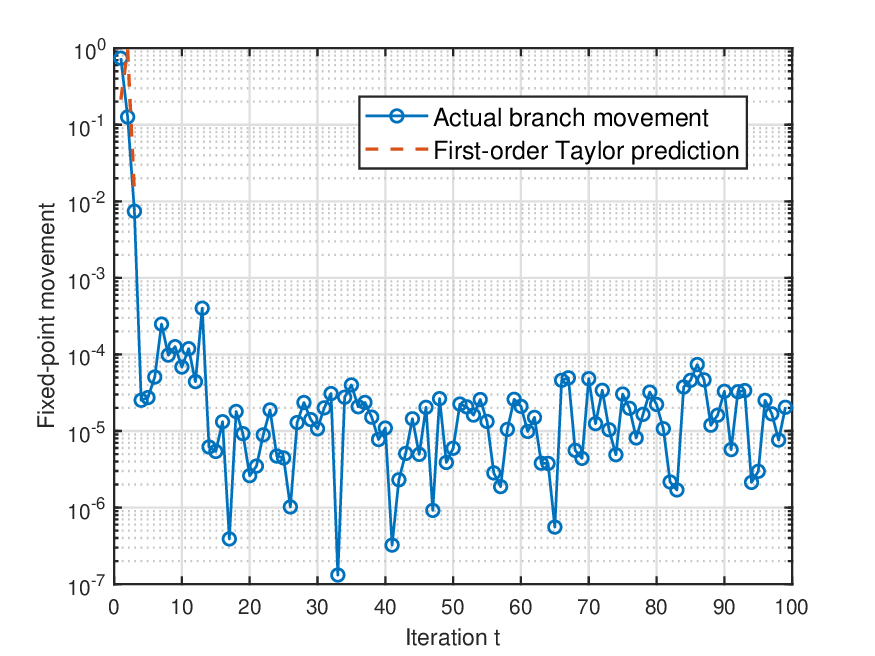}}
	\caption{The temperature-dependent fixed-point and
		Jacobian analyses during the annealing process with $\beta=0.1$.}
	\label{fig:temperature_dynamics2}
\end{figure*}

A similar relationship can be established for the fixed-point
displacement. In the high-temperature region, the Hessian matrix can be
approximated as
\begin{align}
	\mathbf{H}_{T_t}^{\star}
	\approx
	\mathbf{S}_{\rm R}
	+
	2\sigma_n^2T_t\mathbf{I},
\end{align}
such that
\begin{align}
	\lambda_{\min}
	\left(
	\mathbf{H}_{T_t}^{\star}
	\right)
	\approx
	\lambda_{\min}(\mathbf{S})
	+
	2\sigma_n^2T_t
	=
	C_t.
	\label{eq:hessian_min_approximation}
\end{align}
Consequently, as $C_t$ approaches zero from above,
$\|(\mathbf{H}_{T_t}^{\star})^{-1}\|_2$ increases, making the instantaneous
fixed point more sensitive to temperature variations. According to
(\ref{eq:fixed_point_increment}), the turning point of
$\|\boldsymbol{\delta}_t\|_2$ also appears near the zero crossing of $C_t$.

\emph{2) Low-temperature region with $C_t<0$:}
After the temperature crosses the critical region, along the
fixed-point branch being tracked, the posterior-mean components gradually
approach the QPSK constellation levels. For components away from the
decision boundaries, we have
$
	\operatorname{sech}^2
	\left(u_{t,i}^{\star}\right)
	\rightarrow 0.
$
It follows that $\mathbf{D}_t$ gradually decreases and
\begin{align}
	\mathbf{J}_t
	\rightarrow
	(1-\alpha)\mathbf{I},
	\quad
	\rho(\mathbf{J}_t)
	\rightarrow
	1-\alpha.
\end{align}
At the same time, the diagonal term
$\mathbf{D}_t^{-1}$ increases, thereby reducing
$\|(\mathbf{H}_{T_t}^{\star})^{-1}\|_2$. Consequently, the instantaneous
fixed point becomes progressively less sensitive to temperature variations,
and $\|\boldsymbol{\delta}_t\|_2$ eventually approaches zero.

The high-temperature region to the left of these transition points contributes little to the final hard-decision
solution. In this region, the posterior distribution is excessively smooth
and the target fixed point moves rapidly, requiring additional iterations
for the detector to track it. Therefore, the duration of this region should
be minimized. 
This observation motivates initializing the detector near the boundary
between the two regions. Define the critical temperature $T_{\rm c}$ by
\begin{align}
	C_{T_{\rm c}}
	=
	\lambda_{\min}(\mathbf{S})
	+
	2\sigma_n^2T_{\rm c}
	=
	0.
\end{align}
The resulting critical temperature is
\begin{align}
	T_{\rm c}
	=
	-\frac{
		\lambda_{\min}(\mathbf{S})
	}{
		2\sigma_n^2
	}.
	\label{eq:critical_temperature}
\end{align}
Accordingly, the initial temperature is selected as
\begin{align}
	T_0
	=
	\max\left\{
	1,\,
	T_{\rm c}
	\right\}.
	\label{eq:initial_temperature_design}
\end{align}

\section{Simulation Results}
\label{Simulation Results}
In this section, we evaluate the effectiveness of the proposed detector
through Monte Carlo simulations. The simulation parameters are summarized
in Table~\ref{tab:sim_params}. The Doppler shift of each propagation path
is independently generated within
$[-\alpha_{\max},\alpha_{\max}]$. Unless otherwise specified, perfect CSI
is assumed at the receiver to isolate the performance of the detection
algorithm from channel-estimation errors.

\begin{table}
	\caption{Simulation Parameters}
	\label{tab:sim_params}
	\centering
	\begin{tabular}{|c|c|}
		\hline
		\textbf{Parameter} & \textbf{Value} \\
		\hline
		Subcarrier number & $N = 128$ \\
		\hline
		Carrier frequency & $f_c = 5.8~\text{GHz}$ \\
		\hline
		Subcarrier bandwidth & $\Delta f = 15~\text{kHz}$ \\
		\hline
		Modulation order & $M = 4$ \\
		\hline
		Total number of paths & $P = 4$ \\
		\hline
		Maximum normalized Doppler & $\alpha_{\max} = 1$ \\
		\hline
		Maximum normalized delay & $l_{\max} = 3$ \\
		\hline
		Channel coefficients & $h_i \sim \mathcal{CN}(0, 1/P)$ \\
		\hline
	\end{tabular}
\end{table}

\subsection{The Parameter Design of the Annealed IC-based Detector}
In this subsection, we investigate the effects of the principal algorithmic parameters
on the convergence behavior of the proposed detector.

We first investigate the influence of the cooling factor $\beta$.
Fig.~\ref{fig:temperature_dynamics} presents the results for $\beta=0.9$,
whereas Fig.~\ref{fig:temperature_dynamics2} considers an aggressive
cooling schedule with $\beta=0.1$. For $\beta=0.9$, $C_t$ varies smoothly,
$\rho(\mathbf{J}_t)$ eventually approaches $1-\alpha$, and the actual
fixed-point displacement closely matches its first-order prediction,
indicating successful tracking of the moving fixed point.
By contrast, for $\beta=0.1$, $C_t$ changes abruptly from positive to
negative, while $\rho(\mathbf{J}_t)$ remains close to or briefly exceeds
unity. This behavior
indicates transient local instability followed by weak contraction, such
that the tracking error cannot be efficiently attenuated. Since $\|\boldsymbol{\delta}_t\|_2$ scales approximately with
$(1-\beta)T_t$, rapid cooling causes a large fixed-point displacement and
invalidates the first-order approximation, resulting in persistent
tracking fluctuations. Therefore, $\beta$ should be sufficiently close to
one to ensure reliable fixed-point tracking.

\begin{figure}
	\centering
	\includegraphics[width=0.5\textwidth]{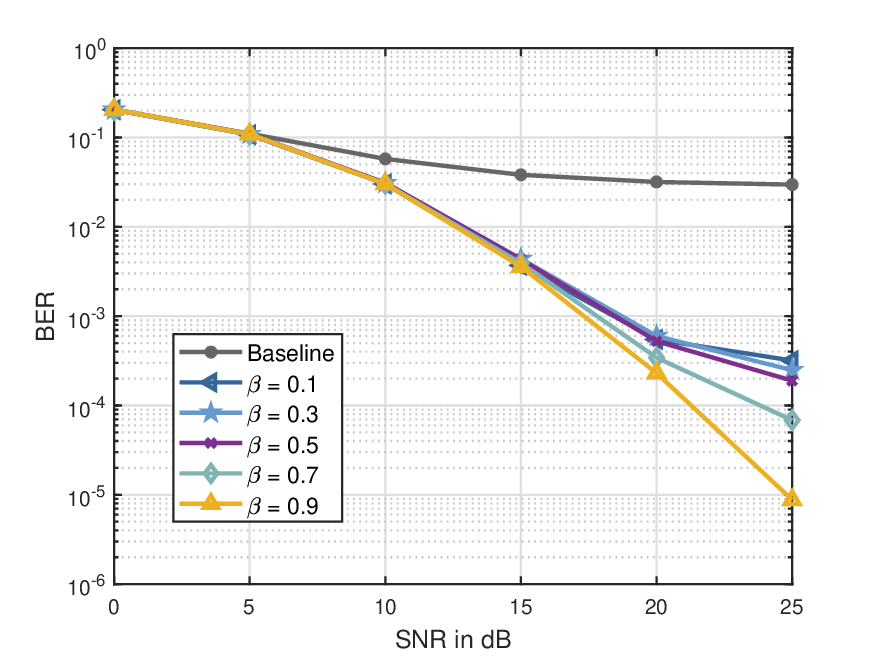}
	\caption{BER performance of the conventional IC detector and the proposed
		annealed IC detector with different cooling factors.}\label{fig:ber_beta}
\end{figure}

Fig.~\ref{fig:ber_beta} compares the BER performance of the conventional
IC detector, denoted as the baseline, with that of the proposed annealed
IC detector under different cooling factors. At low SNRs, all detectors
exhibit similar performance because the detection error is dominated by
additive noise. As the SNR increases, however, the baseline detector
gradually encounters a pronounced error floor. This behavior is consistent
with the preceding analysis: the increasingly sharp posterior distribution
causes the conventional IC detector to behave as a hard-decision dynamical
system, which may become trapped in an erroneous fixed point or limit cycle.
By introducing the annealing mechanism, the proposed detector effectively
removes the high-SNR error floor for all considered values of $\beta$.
Moreover, the BER performance generally improves as $\beta$ increases.
A larger $\beta$ produces a smoother temperature transition and reduces
the displacement between consecutive fixed points, allowing the detector
to track the desired fixed-point branch more reliably. 
These results confirm the effectiveness of the annealing mechanism and
support the use of a relatively slow cooling schedule.

\begin{figure*}[t]
	\centering
	\subfigure[]{
		\label{fig:convexity_margin3}
		\includegraphics[width=0.32\textwidth]{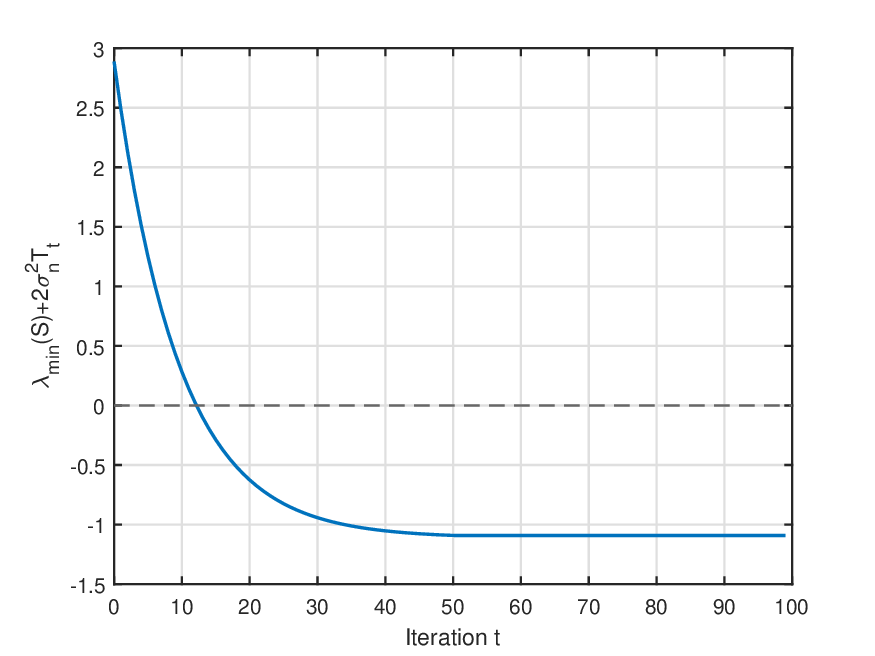}}
	\hfill
	\subfigure[]{
		\label{fig:jacobian_radius3}
		\includegraphics[width=0.32\textwidth]{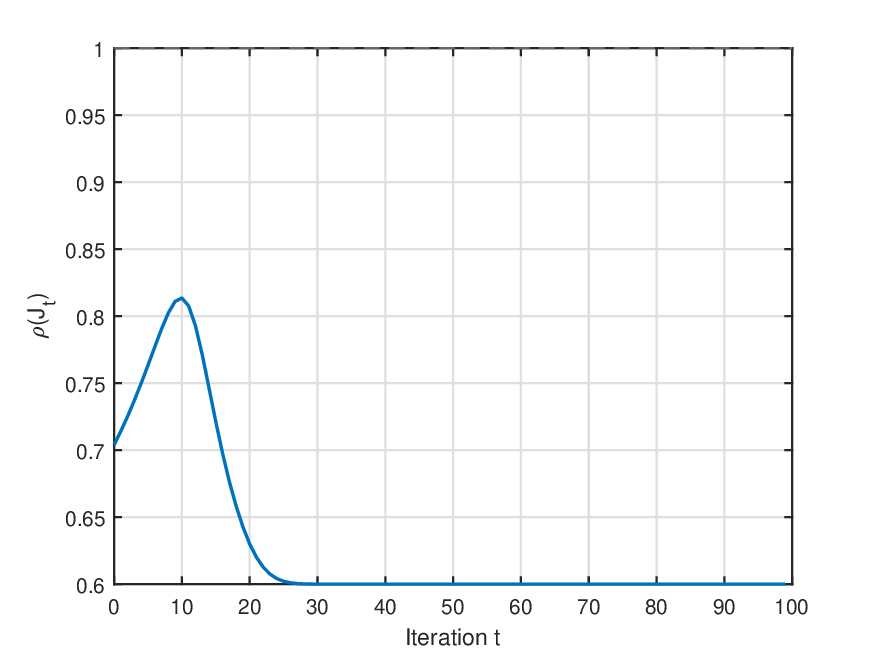}}
	\hfill
	\subfigure[]{
		\label{fig:fixed_point_displacement3}
		\includegraphics[width=0.32\textwidth]{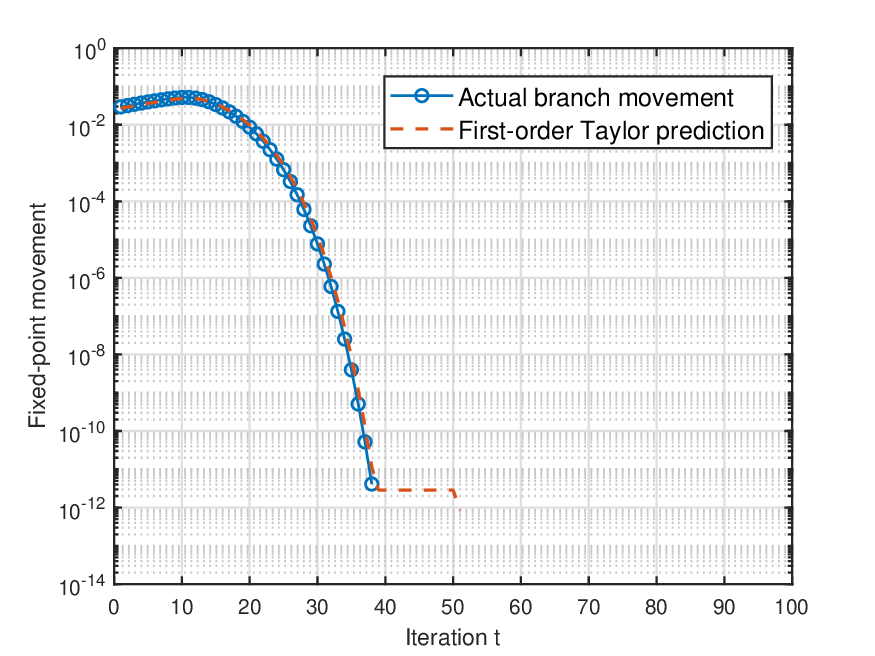}}
	\hfill
	\subfigure[]{
		\label{fig:convexity_margin4}
		\includegraphics[width=0.32\textwidth]{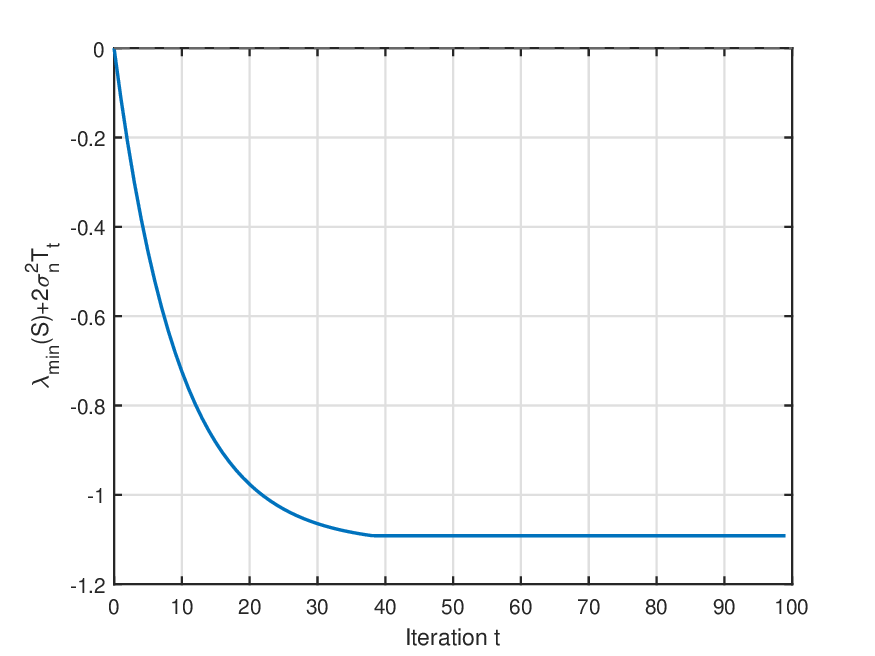}}
	\hfill
	\subfigure[]{
		\label{fig:jacobian_radius4}
		\includegraphics[width=0.32\textwidth]{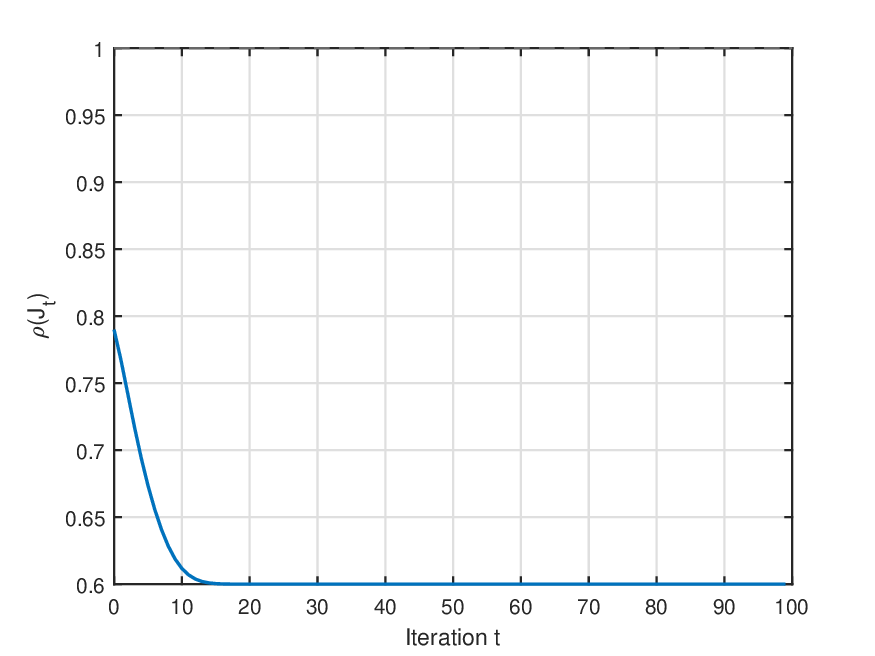}}
	\hfill
	\subfigure[]{
		\label{fig:fixed_point_displacement4}
		\includegraphics[width=0.32\textwidth]{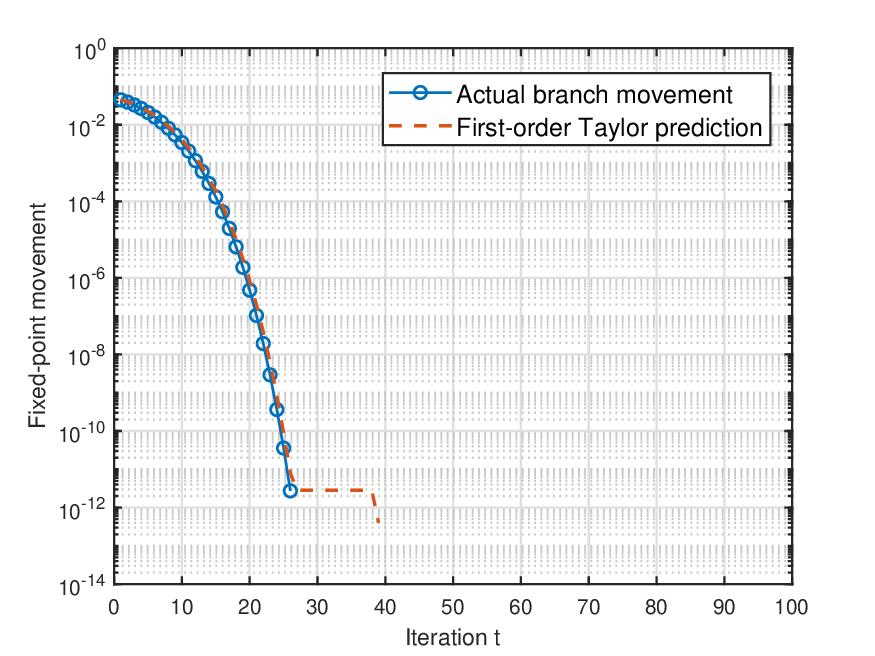}}
	\caption{Effect of the initial-temperature setting on the
		the annealed IC detector:
		(a)--(c) $T_0=6T_{\rm c}$ and (d)--(f) $T_0=T_{\rm c}$.}
	\label{fig:initial_temperature}
\end{figure*}

We then investigate the influence of the initial temperature $T_0$.
The upper row of Fig.~\ref{fig:initial_temperature} corresponds to an
initial temperature $T_0=6T_{\rm c}$, i.e.,
$C_0>0$, whereas the lower row adopts the critical initialization
$T_0=T_{\rm c}$, such that $C_0=0$.
As shown in Figs.~\ref{fig:initial_temperature}(a)--(c), when
$T_0>T_{\rm c}$, the detector first experiences a relatively long
high-temperature stage. During this stage, $\rho(\mathbf{J}_t)$ initially
increases toward its turning point, while the fixed point undergoes
considerable movement. 
By setting $T_0=T_{\rm c}$, the unnecessary high-temperature stage is
effectively removed, as shown in
Figs.~\ref{fig:initial_temperature}(d)--(f). In this case,
$\rho(\mathbf{J}_t)$ decreases directly toward $1-\alpha$, and the
fixed-point displacement decays from the beginning of the iteration.
Consequently, the fixed-point movement reaches a negligible level
approximately ten iterations earlier than that obtained with
$T_0>T_{\rm c}$.
These results demonstrate that initializing the temperature at the
critical boundary can reduce the required number of iterations without
introducing an excessively abrupt transition between fixed-point branches.

\begin{figure}
	\centering
	\includegraphics[width=0.5\textwidth]{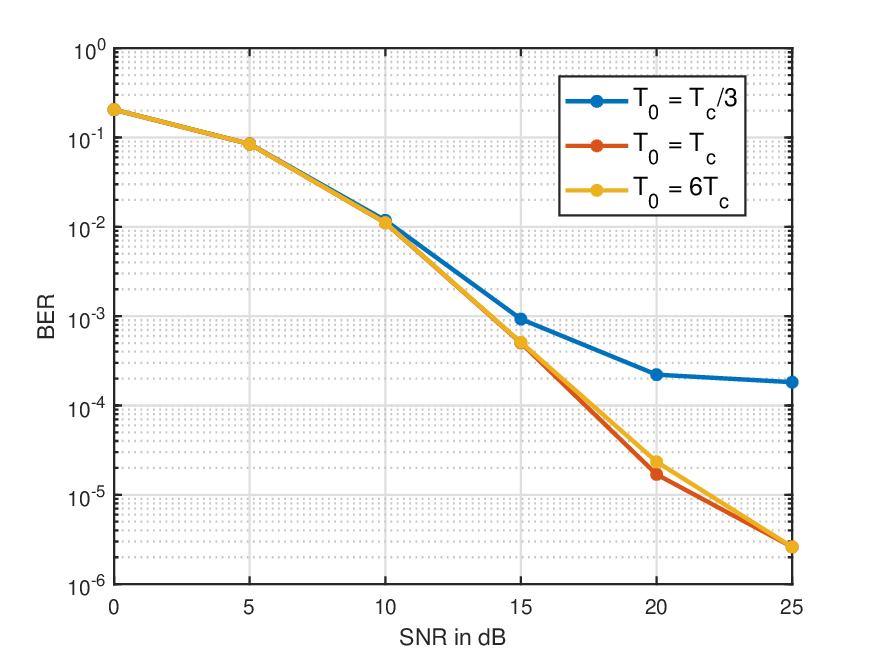}
	\caption{BER performance of the proposed annealed IC detector under
		different initial-temperature settings.}\label{fig:ber_initial_temperature}
\end{figure}

Fig.~\ref{fig:ber_initial_temperature} compares the BER performance under
different initial-temperature settings. At low SNRs, all three settings
exhibit nearly identical performance because the detection error is
dominated by additive noise. Their performance differences become
pronounced as the SNR increases. When $T_0=T_{\rm c}/3$, the temperature
enters the low-temperature regime prematurely, causing the posterior
distribution to sharpen rapidly and restricting the detector's exploration
to nearly discrete constellation decisions. Consequently, the detector
behaves similarly to the conventional IC detector and
exhibits a pronounced high-SNR error floor.
By contrast, both $T_0=T_{\rm c}$ and $T_0=6T_{\rm c}$ effectively eliminate
the error floor and achieve nearly identical BER performance. This result
indicates that increasing the initial temperature beyond $T_{\rm c}$
provides little additional detection gain. However, as demonstrated in
Fig.~\ref{fig:initial_temperature}, an excessively high initial temperature
prolongs the high-temperature tracking stage and increases the required
number of iterations. Therefore, setting $T_0$ near $T_{\rm c}$ achieves
the best tradeoff between detection reliability and convergence speed.

\begin{figure}
	\centering
	\includegraphics[width=0.5\textwidth]{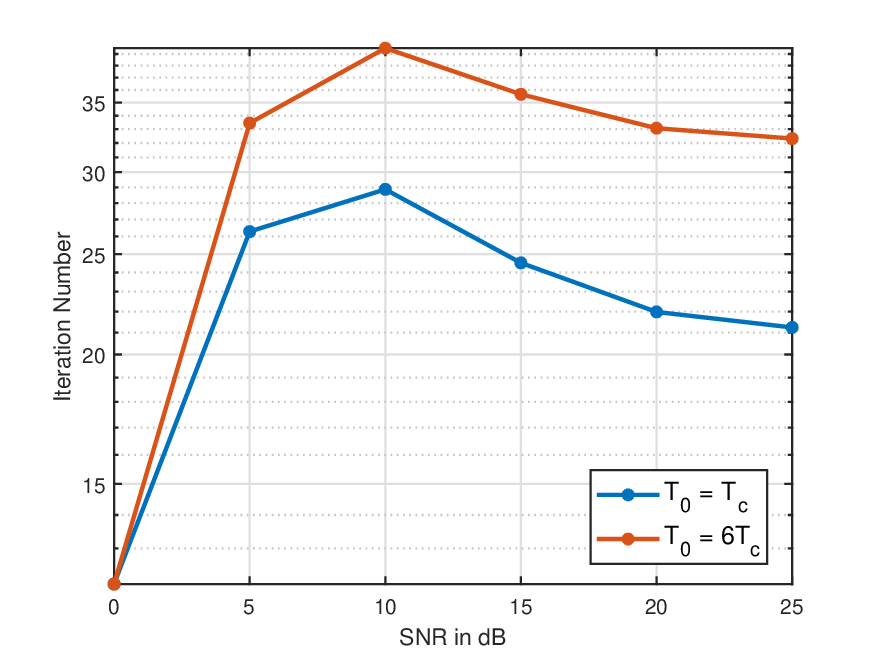}
	\caption{Average number of iterations required for convergence under
		different initial-temperature settings.}\label{fig:average_iteration}
\end{figure}

Fig.~\ref{fig:average_iteration} compares the average number of iterations
required for convergence under $T_0=T_{\rm c}$ and $T_0=6T_{\rm c}$. At
very low SNR, the two settings require similar numbers of iterations
because the convergence behavior is dominated by noise. As the SNR
increases, however, initializing at $T_{\rm c}$ consistently converges
faster than initializing at $6T_{\rm c}$.
This reduction can be explained by the additional high-temperature stage
introduced by $T_0=6T_{\rm c}$. 
These extra iterations mainly track the rapidly moving high-temperature fixed
point and provide little improvement in the final detection performance.
Consequently, setting $T_0=T_{\rm c}$ reduces the average iteration count
by approximately $20\%$--$35\%$ over the considered SNR range.

\subsection{Performance Comparison with Different Detectors}
In this subsection, we compare the BER performance and computational
complexity of the proposed detector with those of several representative
detection algorithms. The considered benchmarks include the linear MMSE
detector, the Gaussian-EP detector~\cite{Céspedes2014TCom}, and the Gaussian symbol approximation (GS)-MP
detector~\cite{Conti2026TWC}. 

\begin{figure}
	\centering
	\includegraphics[width=0.5\textwidth]{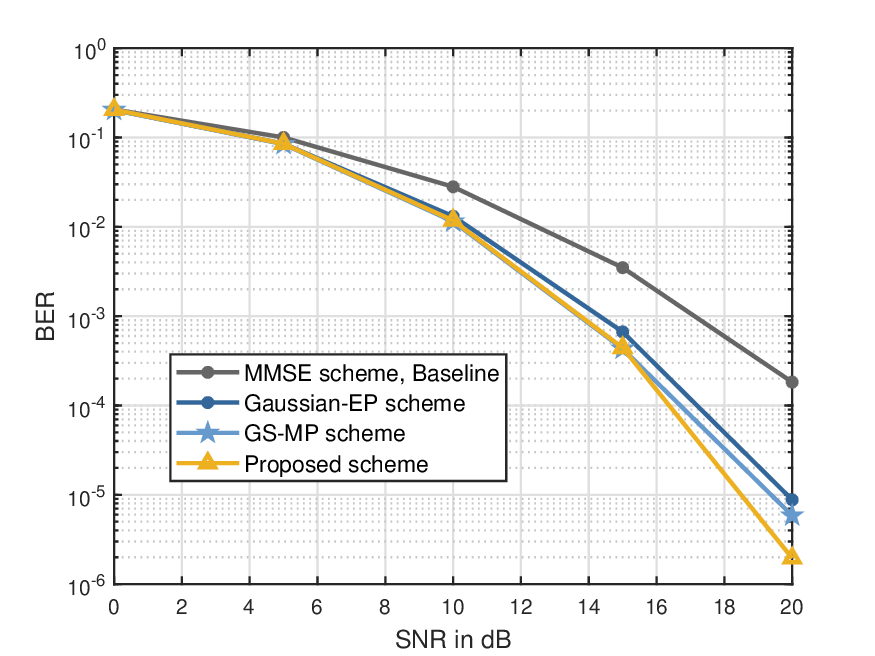}
	\caption{BER performance comparison of different detection algorithms.}\label{fig:detector_ber}
\end{figure}

The BER comparison is presented in first
Fig.~\ref{fig:detector_ber}.
As shown in Fig.~\ref{fig:detector_ber}, all the considered detectors
exhibit comparable BER performance in the low-SNR regime, where additive
noise dominates the detection error. As the SNR increases, the performance
advantage of the proposed detector becomes increasingly pronounced. The
linear MMSE detector exhibits the slowest BER reduction because its
one-shot linear processing cannot sufficiently explore the multi-path diversity gain.
The Gaussian-EP and GS-MP detectors significantly outperform the MMSE
detector through iterative interference refinement. Nevertheless, the
proposed detector consistently achieves the lowest BER in the moderate-
and high-SNR regimes. 
This performance gain can be attributed to the temperature-controlled
posterior update, which preserves soft exploration during the early
iterations and gradually sharpens the symbol estimates as the iteration
proceeds. 

\begin{figure}
	\centering
	\includegraphics[width=0.5\textwidth]{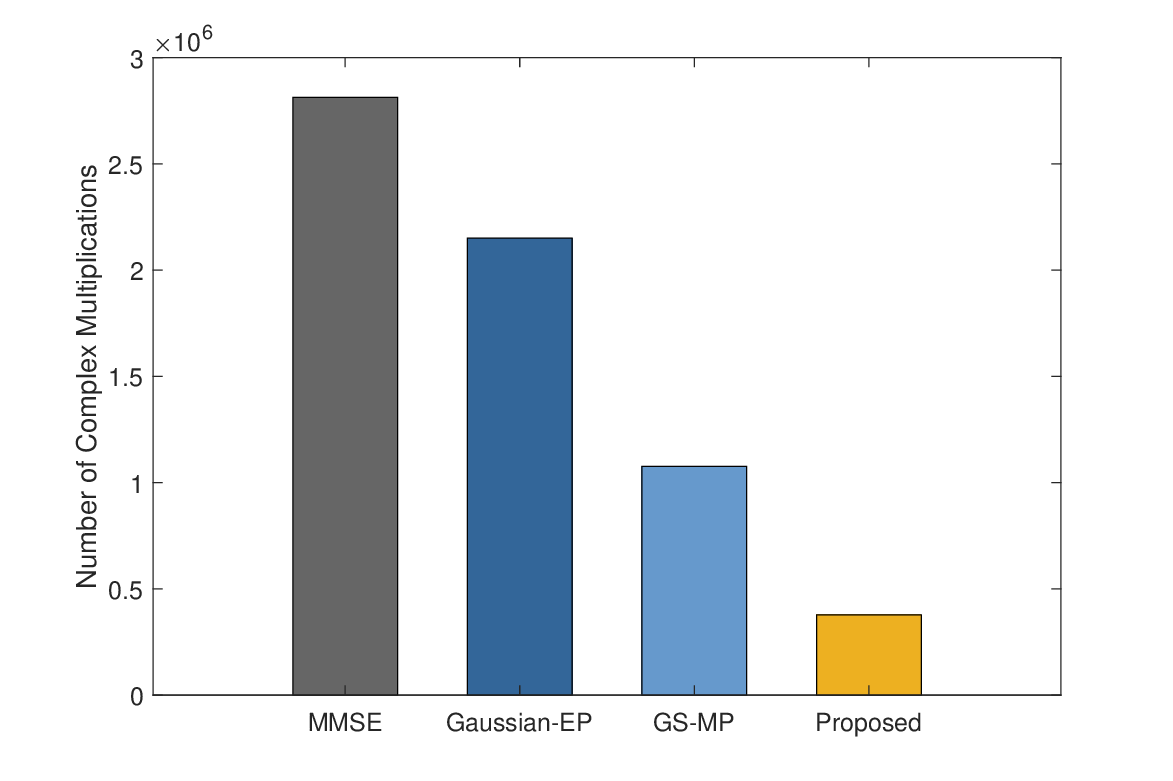}
	\caption{Computational complexity comparison of different detection
		algorithms in terms of complex multiplications.}\label{fig:complexity_comparison}
\end{figure}

Fig.~\ref{fig:complexity_comparison} compares the computational complexity
of different detectors in terms of the number of complex multiplications
under the same system configuration. The MMSE detector incurs the highest
complexity, requiring approximately $2.8\times10^6$ complex
multiplications, primarily due to the high-dimensional matrix inversion.
The Gaussian-EP detector reduces the complexity to approximately
$2.15\times10^6$, and the GS-MP detector requires approximately $1.08\times10^6$ complex
multiplications. In comparison, the proposed detector requires only
approximately $0.37\times10^6$ complex multiplications. This corresponds
to complexity reductions of approximately $87\%$, $83\%$, and $66\%$
relative to the MMSE, Gaussian-EP, and GS-MP detectors, respectively.
The complexity advantage of the proposed detector mainly arises from its avoidance of explicit interference-statistic calculations. The proposed IC detector employs the annealing temperature to implicitly characterize the uncertainty caused by residual interference and directly controls the softness of the posterior distribution.

\section{Conclusion}
\label{Conclusion}
This paper presented a low-complexity symbol detection framework for AFDM
based on the Ungerboeck observation model. We first investigated the
high-SNR performance degradation of conventional IC detection and
demonstrated that posterior sharpening transforms the soft detector into
a finite-state hard-decision dynamical system, whose trajectory may become
trapped in an erroneous fixed point or finite-period limit cycle. To
address this issue, an annealed IC detector incorporating
temperature-controlled posterior updates and damping was proposed.
A temperature-dependent potential function and a time-varying Jacobian
analysis were further developed to guide the selection of the initial
temperature and cooling factor. Simulation results validated the
theoretical analysis and demonstrated that the proposed detector achieves
superior detection performance with substantially lower computational
complexity, providing an effective low-complexity detection solution for
AFDM systems.

\bibliographystyle{IEEEtran}%
\bibliography{bib/bibfile}

\vfill

\end{document}